\documentclass[twocolumn,showpacs,floatfix,eqsecnum,prb]{revtex4}

\usepackage[dvips]{epsfig}

\usepackage{float}

\begin{document}

\title{Topological Insulators with Inversion Symmetry}

\author{Liang Fu and C.L. Kane}
\affiliation{Dept. of Physics and Astronomy, University of
Pennsylvania, Philadelphia, PA 19104}

\begin{abstract}
Topological insulators are materials with a bulk excitation gap
generated by the spin orbit interaction, and which are different from
conventional insulators.  This distinction is characterized by $Z_2$
topological invariants, which characterize the groundstate.  In two
dimensions there is a single $Z_2$ invariant which distinguishes the
ordinary insulator from the quantum spin Hall phase.  In three
dimensions there are four $Z_2$ invariants, which distinguish the
ordinary insulator from ``weak" and ``strong" topological insulators.
These phases are characterized by the presence of gapless surface
(or edge) states.  In the 2D quantum spin Hall phase and the 3D strong
topological insulator these states are robust and
are insensitive to weak disorder and
interactions.  In this paper we show that the
presence of inversion symmetry greatly simplifies the problem of
evaluating the $Z_2$ invariants.  We show that the invariants can be
determined from the knowledge of the parity of the occupied Bloch
wavefunctions at the time reversal invariant points in the Brillouin
zone.  Using this approach, we predict a number of specific materials
are strong topological insulators, including the semiconducting alloy
Bi$_{1-x}$ Sb$_x$ as well as $\alpha$-Sn and HgTe under uniaxial
strain.  This paper also includes an expanded discussion of our formulation of
the topological insulators in both two and three dimensions, as well
as implications for experiments.

\end{abstract}

 \maketitle

\section{Introduction}

In elementary solid state physics textbooks an insulator is described
as a material with an energy gap separating filled and empty energy
bands.  A more sophisticated definition of an insulator is that of a
material for which all electronic phenomena are {\it local}\cite{kohn}.  Such a
material is insensitive to boundary conditions, so that in a multiply
connected sample, such as a ring, there is exponentially small sensitivity to magnetic
flux threading the holes.   The simplest atomic insulator,
in which all electrons are tightly bound to atoms in closed shells,
clearly satisfies both properties.  Ionic and covalent
insulators do too.  These band insulators are topologically
equivalent in the sense that
the Hamiltonian can be adiabatically transformed into an atomic
insulator without going through any phase transitions.  With regards
to their low energy electronic behavior, conventional insulators are
equivalent to atomic insulators.

The existence of a bulk energy gap does not guarantee the
insensitivity to boundary conditions, and there exist phases with
bulk gaps, which are topologically distinct.
In addition to exotic strongly correlated
phases\cite{wen,demler}, this fact arises even for non
interacting electrons described within band theory.  The simplest
example is the integer quantum Hall effect (IQHE).  In two dimensions,
a magnetic field introduces a cyclotron gap between Landau levels,
which may be viewed as energy bands in the magnetic Brillouin zone.
This phase can exist even without Landau levels in the absence of a
uniform magnetic field\cite{haldane},
though a necessary condition is that time reversal
symmetry be broken.  Based on the bandstructure alone it is
difficult to tell the difference between the IQHE state and a band insulator.
The distinction between the two is a topological property of the
occupied bands which is encoded into the Chern integer introduced
by Thouless et al. \cite{tknn}.
Three dimensional generalizations of the IQHE state, which may be
viewed as layered 2D states, are indexed by a triad of Chern
integers\cite{kohmoto}.  A hallmark of the IQHE phases, which is intimately related to
their topology, is the existence of
gapless chiral edge states which are robust in the presence of
disorder\cite{halperin,hatsugai1}.

Recently, new topological
insulating phases for systems with time reversal symmetry have been discovered
\cite{km1,km2,bernevig,roy1,moore,roy2,fkm}.
In two
dimensions, the quantum spin Hall phase is distinguished from a
band insulator by a single $Z_2$ invariant\cite{km2}.  This phase exhibits
gapless spin-filtered edge states, which allow for dissipationless
transport of charge and spin at zero temperature, and are
protected from weak disorder and interactions by time reversal
symmetry.  In three dimensions a time reversal invariant
bandstructure is characterized by four $Z_2$ invariants\cite{moore,roy2,fkm}.
Three of
the invariants rely on the translational symmetry of the lattice
and are not robust in the presence of disorder, leading to ``weak
topological insulators".  The fourth
invariant, however, is robust and distinguishes the ``strong
topological insulator" (STI).

Nontrivial $Z_2$ invariants imply the existence of gapless surface
states.  In particular, in the STI phase, the surface states form
a two dimensional ``topological metal", in which the Fermi arc
encloses an {\it odd} number of Dirac points\cite{fkm}.  This
leads to a quantized Berry's phase of $\pi$ acquired by an
electron circling the surface Fermi arc, which does not change
under continuous perturbations\cite{haldane2,suzuura}.  The $\pi$
Berry's phase also signifies that with disorder the surface states
are in the symplectic universality class, and exhibit
antilocalization\cite{hikami}.  Thus, the metallic surface states
form a unique phase, which can not be realized in a conventional
two dimensional electron system for which Dirac points must come
in pairs\cite{nielson}.

The purpose of this paper is twofold.  First, we will explain the
formulation of the $Z_2$ invariants in somewhat more detail
than in Ref. \onlinecite{fkm}, emphasizing the physical meaning
of the invariants and their relation to the surface states.
Second, we will develop a new
method for evaluating the $Z_2$ invariants in systems which have
inversion symmetry.  This allows us to predict a number of
specific materials to be strong topological insulators.

Most insulators are conventional insulators.  In order to find topological
insulators experimentally it is necessary to
develop criteria for recognizing them from their
bulk band structure.  Clearly, a necessary condition is the existence
of a bulk energy gap which owes its existence to the spin-orbit
interaction.  However, evaluating the $Z_2$ invariants for a given
band structure is in general a difficult problem.  To date three
general approaches have been used.

(1)  One powerful approach is to exploit adiabatic continuity to a
Hamiltonian which has extra
symmetry.  We used this method to identify the quantum spin Hall phase in
graphene\cite{km1,km2}
by arguing that the Hamiltonian for graphene is adiabatically
connected to a Hamiltonian in which the spin $S_z$ is conserved.
With this extra conservation law the system can be characterized by a
spin {\it Chern number}, which describes the
quantized spin Hall conductivity\cite{sheng,hatsugai}.
The $Z_2$ invariant can then be identified
with the {\it parity} of the spin Chern number.  In the presence of
$S_z$ non conserving terms (which are inevitably present) the spin
Chern number loses its meaning\cite{fk1}.  However, the $Z_2$ invariant retains
its value and characterizes the quantum spin Hall phase.

Adiabatic continuity can also be used to establish a material is a
band insulator if an adiabatic path can be found which connects
the material to an ``atomic" limit.  Moreover, it can be argued
that the $Z_2$ invariant {\it changes} at an appropriate quantum
phase transition, where the bulk energy gap goes to
zero\cite{roy1,roy2}.  In general, this approach require a
continuous path be found which connects the Hamiltonian in
question to a known phase.

(2) It is also possible to evaluate the $Z_2$ invariant directly with the
knowledge of the Bloch wavefunctions for the occupied energy bands.
In Ref. \onlinecite{fk1} we established a formula for the invariant in terms of a
Brillouin zone integral.  This is analogous to
the calculation of the Chern number as an integral of the gauge invariant
Berry's curvature\cite{tknn,kohmoto2}.  However, unlike the Chern invariant,
the integral for the $Z_2$ invariant also involves the Berry's
{\it potential}, and requires a gauge in
which the wavefunctions are {\it globally} continuous.  Since time reversal
symmetry requires the Chern invariant to vanish, a globally
continuous gauge is guaranteed to exist.  However, finding a
continuous gauge is not always simple.

(3)  A third approach
is to characterize the zeros of Pfaffian function introduced Ref. \onlinecite{km2}.
Though the Pfaffian is not gauge invariant, its zeros can be
determined without specifying a continuous gauge.  While this
approach is tedious (especially in three dimensions) it has been
successfully implemented by Murakami\cite{murakami1} to show that 2 dimensional
bismuth bilayers realize a quantum spin Hall phase.

In this paper we will show that the presence of inversion symmetry
greatly simplifies the problem of identifying the $Z_2$ invariants.
We show that the invariants can be determined from the knowledge of
the parity of the occupied band eigenstates at the eight (or four in
two dimensions) time reversal invariant momenta $\Gamma_i$ in the Brillouin zone.
Specifically, we will show that the $Z_2$ invariants are determined
by the quantities
\begin{equation}
\delta_i = \prod_{m=1}^N \xi_{2m}(\Gamma_i).
\label{eq1}
\end{equation}
Here $\xi_{2m}(\Gamma_i)=\pm 1$ is the parity eigenvalue of the
$2m'th$ occupied energy band at $\Gamma_i$, which shares the same
eigenvalue $\xi_{2m} = \xi_{2m-1}$ with its Kramers degenerate
partner.  The product involves the $2N$ occupied bands. The $Z_2$
invariant $\nu=0, 1$ which distinguishes the quantum spin Hall phase in two
dimensions and the strong topological insulator in three
dimensions is then given by the product of all the $\delta_i$'s,
\begin{equation}
(-1)^\nu = \prod_i \delta_i.
\label{eq2}
\end{equation}
The other three ``weak" topological invariants in three dimensions
also are determined by $\delta_i$.
Since the parity eigenvalues $\xi_n(\Gamma_i)$ are tabulated in the
band theory literature
this allows us to identify inversion symmetric
topological  insulating materials.  Moreover, exploiting adiabatic continuity
allows us to identify topological insulators which don't have
inversion symmetry, but are adiabatically connected to materials
which have inversion symmetry.

Applying the above approach, we predict that the following narrow
gap semiconductors are strong topological insulators: (1) the
alloy Bi$_{1-x}$Sb$_x$, which is semiconducting for $.07<x<.22$.
(2) $\alpha-$Sn and HgTe under uniaxial strain, (3) The alloy
Pb$_{1-x}$Sn$_x$Te under uniaxial strain for $x \sim x_c$ in the
vicinity of the band inversion transition. The materials (2-3)
were suggested by Murakami, Nagaosa and Zhang\cite{murakami2} as
candidates for spin Hall insulators.  Those authors argued that
those materials share a large spin-Hall conductivity, as
calculated by a Kubo formula.  Our analysis of these materials
is rather different, and we will show that PbTe is a
conventional insulator, despite its large spin-Hall conductivity,
while strained $\alpha-$Sn and HgTe are topological insulators.

In section II we will present an expanded discussion of our
formulation of the $Z_2$ invariants.  Then, in section III, we will
derive Eqs. (1.1,1.2) for problems with inversion symmetry. In
section IV we will apply our method to a class of four band tight
binding models, which includes the graphene model as well as the
3D model introduced in Ref. \onlinecite{fkm}. In section V we will
apply (1.1,1.2) to deduce the $Z_2$ invariants of several real
materials based on their known band structures.  Readers
uninterested in the technical details can skip directly to section
V to read about these applications. Finally, in section VI we will
conclude with a brief discussion of the experimental implications
for the topological insulating phases.

\section{$Z_2$ Invariants in Two and Three Dimensions}

In this section, we will review our formulation of the topological insulating
phases.  We begin in IIA by defining the time reversal polarization.
In IIB, we develop the $Z_2$ characterization of a bandstructure as a
topological property of the occupied Bloch wavefunctions.  In IIC we show
how the $Z_2$ invariants determine the surface state spectrum.  In
IIC, we consider a more general formulation of the $Z_2$ invariant as
a sensitivity of a bulk crystal to boundary conditions.

\subsection{The Time Reversal Polarization}

In Ref. \onlinecite{fk1} we introduced the concept of the {\it
time reversal polarization}, in the same spirit as charge
polarization. This allows for an interpretation of the $Z_2$
invariants using a Laughlin type gedanken experiment on a
cylinder\cite{laughlin}. In the ordinary quantum Hall effect, the
Chern invariant can be interpreted in a similar way.  To motivate
the time reversal polarization we therefore begin with a
discussion of the charge polarization.

The charge polarization determines the surface charges present in a
finite system.  Since electrons may be added or removed from a
surface, the charge polarization is only defined modulo an
integer\cite{blount,zak,kingsmith,resta}.
However, {\it changes} in the charge polarization induced
by adiabatic changes in the Hamiltonian are well defined.  In
Laughlin's gedanken experiment for the integer quantum Hall effect a
quantum of magnetic flux $h/e$ is adiabatically inserted in a cylindrical
quantum Hall sample at filling $\nu = N$.  The resulting transfer of
$N$ electrons from one end of the cylinder to the other can be interpreted as a
change in the charge polarization of the cylinder.  The
Chern invariant, which characterizes the integer quantum Hall state
precisely characterizes this quantized change in charge polarization.

The time reversal polarization is a $Z_2$ quantity, which
characterizes the presence or absence
of a Kramers degeneracy associated with a surface.  Like the charge
polarization, its value can be changed by adding an extra electron to
the surface.  Thus by itself, the time reversal polarization is not
meaningful.  However, {\it changes} in the time reversal polarization
due to adiabatic changes in the bulk Hamiltonian {\it are} well defined.
Specifically the change in the time reversal polarization when
{\it half} a flux quantum $h/2e$ is threaded through a cylinder defines
a $Z_2$ invariant which is analogous to the Chern invariant, and
distinguish topological insulators.

\subsection{$Z_2$ invariants characterizing a bandstructure}

The topological invariant characterizing a two dimensional band
structure may be constructed by imagining a long cylinder whose
axis is parallel to a reciprocal lattice vector ${\bf G}$, and
which has a circumference of a single lattice constant.  Then the
magnetic flux threading the cylinder plays the role of the
circumferential (or ``edge'') crystal momentum $k_x$, with
$\Phi=0$ and $\Phi=h/2e$ corresponding to two ``edge" time
reversal invariant momenta $k_x = \Lambda_1$ and $k_x=\Lambda_2$.
The $Z_2$ invariant characterizes the change in the Kramers
degeneracy at the ends of this 1 dimensional system between $k_x =
\Lambda_1$ and $k_x=\Lambda_2$.

For a three dimensional crystal, imagine a ``generalized cylinder" which is
long in one direction (parallel to ${\bf G}$), but in the other two directions
has a width of one lattice constant with periodic boundary conditions.
While this structure can not be pictured as easily as a regular cylinder, a
distorted (but topologically equivalent) version can be visualized as
a torus with a finite thickness.  This ``corbino donut" is
analogous to the generalized cylinder in the same way the corbino
disk is analogous to the regular cylinder.
The ``long" direction corresponds to the thickness of the
torus, and the two ends correspond to the inner and outer surfaces.
This system can be threaded by two independent magnetic fluxes, and
they correspond to the two components of the momentum perpendicular
to ${\bf G}$.  There are four time reversal invariant surface momenta $\Lambda_a$,
corresponding to the two fluxes being either $0$ or $h/2e$.  The
bandstructure can be characterized by the difference in the time reversal
polarization between any pair.

\begin{figure}
 \centerline{ \epsfig{figure=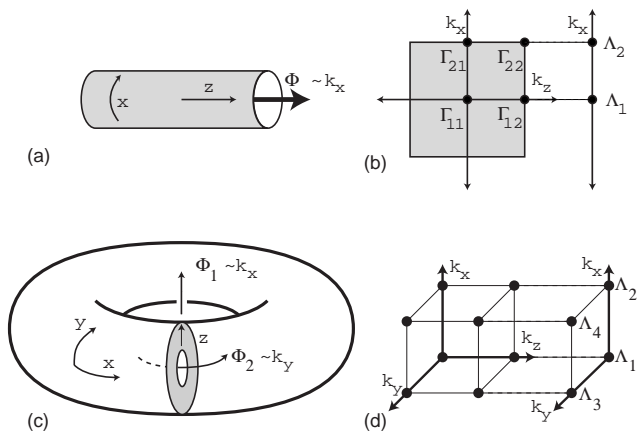,width=3.3in} }
 \caption{(a) A two dimensional cylinder threaded by magnetic flux
 $\Phi$.  When the cylinder has a circumference of a single lattice
 constant $\Phi$ plays the role of the edge crystal momentum $k_x$ in band
 theory.  (b) The time reversal invariant fluxes $\Phi=0$ and $h/2e$
 correspond to edge time reversal invariant momenta $\Lambda_1$ and
 $\Lambda_2$.  $\Lambda_a$ are projections of pairs of the four bulk
  time reversal momenta $\Gamma_{i=(a\mu)}$, which reside in the two
  dimensional Brillouin zone indicated by the shaded region.
  (c)  In 3D the generalized cylinder can be visualized as a
  ``corbino donut", with two fluxes, which correspond to the
  two components of the surface crystal momentum.  (d) The
  four time reversal invariant fluxes $\Phi_1$, $\Phi_2 = 0$, $h/2e$
  correspond to the four two dimensional surface momenta $\Lambda_a$.
  These are projections of pairs of the eight $\Gamma_{i=(a\mu)}$
  that reside in the bulk 3D Brillouin zone.
 }
 \label{cylinderfig}
 \end{figure}

The $Z_2$ invariants can be deduced from the topological structure of
the Bloch wavefunctions of the bulk crystal in the Brillouin zone.
Consider a time reversal invariant
periodic Hamiltonian ${\cal H}$ with $2N$ occupied bands
characterized by Bloch wavefunctions
\begin{equation}
|\psi_{n, {\bf k}}\rangle=e^{i \bf k \cdot {\bf r}}  |u_{n, {\bf
k}}\rangle.
\end{equation}
Here $|u_{n, \bf k}\rangle$ are cell periodic eigenstates of the Bloch
Hamiltonian,
\begin{equation}
H({\bf k})= e^{-i \bf k \cdot r} {\cal H} e^{+i \bf k \cdot r}.
\label{Bloch}
\end{equation}
We require
$|\psi_{n,{\bf k}+{\bf G}}\rangle=|\psi_{n,{\bf
k}}\rangle$ for reciprocal lattice vectors ${\bf G}$, so that
the Brillouin zone in which ${\bf k}$ is defined
is a torus.  This implies $|u_{n,{\bf k}+{\bf G}}\rangle =
e^{-i {\bf G} \cdot {\bf r}}|u_{n,{\bf k}}\rangle$.
Time reversal symmetry implies $[{\cal H},\Theta] = 0$,
where $\Theta = \exp(i\pi S_y)K$ is the time reversal
operator ($S_y$ is spin and $K$ complex conjugation), which
for spin 1/2 particles satisfies $\Theta^2=-1$.
It follows that $H(-{\bf k}) = \Theta H({\bf k})
\Theta^{-1}$.

There are special points ${\bf k} = \Gamma_i$
in the Brillouin zone which are time
reversal invariant and satisfy $-\Gamma_i = \Gamma_i+{\bf G}$ for a
reciprocal lattice vector ${\bf G}$.  There are eight such points in
three dimensions and four in two dimensions.  At these points
$H(\Gamma_i) = \Theta H(\Gamma_i) \Theta^{-1}$, so that the
eigenstates are Kramers degenerate.
In the following it will be useful to use two different notations to label the
distinct $\Gamma_i$.
(1)  The eight (or four) $\Gamma_i$ can be indexed by three (or two)
integers $n_l=0,1$ defined modulo 2, which specify
half a ``mod 2 reciprocal lattice vector",
\begin{equation}
\Gamma_{i=(n_1n_2n_3)} ={1\over 2}\left(n_1 {\bf b}_1 + n_2 {\bf b}_2 \
+ n_3 {\bf b}_3  \right) ,
\label{gnk}
\end{equation}
where ${\bf b}_l$ are primitive reciprocal lattice vectors.
Two mod 2 reciprocal lattice vectors are equivalent if they
differ by twice a reciprocal lattice
vector.  (2) A second notation is useful
when considering a surface perpendicular to reciprocal lattice vector
${\bf G}$.  The surface time reversal invariant momenta $\Lambda_a$ will be
projections of pairs of $\Gamma_i$ which differ by ${\bf G}/2$,
as shown in Fig. \ref{cylinderfig}.
Given ${\bf G}$, we can define $\Gamma_{i=(a\mu)}$, such that
$\Gamma_{a1}-\Gamma_{a2}={\bf G}/2$.

The change in the time reversal polarization
associated with a cylinder oriented along ${\bf G}$ due to changing
the flux between two values corresponding to $\Lambda_a$ and
$\Lambda_b$
can be calculated by a method analogous to the calculation of the
charge polarization as a Berry's phase\cite{blount,zak,kingsmith,resta}.
In Ref. \onlinecite{fk1} we showed that
the result is most simply expressed in terms of the quantities
\begin{equation}
\delta_i =
{\sqrt{{\rm det}[w(\Gamma_i)]}\over
{{\rm Pf}[w(\Gamma_i)]}} = \pm 1,
\label{Z2}
\end{equation}
where
$w$ is the $2N\times 2N$ unitary matrix defined by
\begin{equation}
w_{mn}({\bf k}) \equiv \langle u_{m-{\bf k}}|\Theta|u_{n{\bf k}}
\rangle. \label{w}
\end{equation}
Since $\langle\Theta a|\Theta b\rangle = \langle b|a\rangle$ and
$\Theta^2=-1$ the matrix
$w({\bf k})$ is antisymmetric for ${\bf k}=\Gamma_i$, allowing for
the definition of the Pfaffian, which satisfies ${\rm det}[w] = {\rm
Pf}[w]^2$.
Given a surface ${\bf G}$ the time
reversal polarization associated with surface momentum $\Lambda_a$
was found to be
\begin{equation}
\pi_a = \delta_{a1}\delta_{a2}.
\end{equation}
This formula is analogous to the expression for the charge
polarization as a Berry's phase
along a closed cycle in momentum space\cite{zak}.

Because of the square root, the sign of $\delta_i$ is ambiguous.
However, since we require $|u_{n,{\bf k}}\rangle$ to be
continuous, $\sqrt{{\rm det}[w({\bf k})]}$ is
defined {\it globally} throughout the Brillouin zone.  The sign
ambiguity then cancels for $\pi_a$.  The existence of the global
square root is guaranteed by the fact that the phase winding of
${\rm det} [w(\bf k)]$ around any closed loop $C$ is zero, so that
$\sqrt{{\rm det}[w({\bf k})]}$ has no branch cuts. When $C$ is
contractable, the vanishing phase winding follows from the
continuity of $|u_{n,{\bf k}}\rangle$.  For non contractable
loops, which can be continuously deformed to satisfy $C=-C$, it
follows from the fact that
 ${\rm det} [w(-{\bf k})]={\rm det} [w(\bf k)]$.

The inherent ambiguity of the time reversal polarization is reflected
in the fact that the $\pi_a$ are not gauge invariant.
For instance, consider a gauge transformation of the form
\begin{equation}
|u_{n,{\bf k}}\rangle \rightarrow
\left\{\begin{array}{rl}
e^{i \theta_{{\bf k}}} |u_{n,{\bf k}}\rangle & {\rm for}\quad n=1 \\
|u_{n,{\bf k}}\rangle & {\rm for}\quad n\ne 1.
\end{array}\right.
\label{newgauge}
\end{equation}
Under this transformation,
\begin{eqnarray}
{\rm det}[w({\bf k})]& \rightarrow &
 {\rm det}[w({\bf k})] e^{-i(\theta_{{\bf k}}+\theta_{{\bf -k}})}, \label{detchange}\\
{\rm Pf}[w(\Gamma_i)]& \rightarrow &
 {\rm Pf}[w(\Gamma_i)] e^{-i\theta_{\Gamma_i}}.
 \end{eqnarray}
If we choose $\theta_{{\bf k}}={\bf k}\cdot {\bf R}$ for a
lattice vector ${\bf R}$ the Brillouin zone periodicity of $u_{n{\bf k}}$ is
 preserved.  From \ref{detchange} it is clear that ${\rm det}[w({\bf k})]$ is unchanged.
 However, if ${\bf G}\cdot{\bf R}=2\pi$, it follows that
 ${\rm Pf}[w(\Gamma_{a1})] {\rm Pf}[w(\Gamma_{a2})]$, and hence
 $\pi_a$,
changes sign.  Since this
 gauge transformation changes the sign of $\pi_a$ for all $a$,
 however, the product $\pi_a\pi_b$, which gives the {\it change} in the
 time reversal polarization between $\Lambda_a$ and $\Lambda_b$ remains invariant.
 In general, the product of any four $\delta_i$'s for which $\Gamma_i$
lie in the same plane is gauge invariant, and defines a topological invariant
 characterizing the band structure.

In two dimensions there are four time reversal invariant momenta
$\Gamma_i$ and a single $Z_2$ invariant, given by
\begin{equation}
(-1)^\nu = \prod_{i=1}^4  \delta_i.
\label{nueq}
\end{equation}
In three dimensions there are 8 time reversal invariant points.
This leads to 4 independent $Z_2$ invariants\cite{moore,roy2,fkm}.
One of these invariants, $\nu_0$, can be expressed as the product over all eight
points,
\begin{equation}
(-1)^{\nu_0} = \prod_{i=1}^8 \delta_i
\label{nu0eq}
\end{equation}
The other three invariants are given by products of four $\delta_i$'s for
which $\Gamma_i$ reside in the same plane.
\begin{equation}
(-1)^{\nu_k} = \prod_{n_k=1;n_{j\ne k}=0,1} \delta_{i=(n_1n_2n_3)}.
\label{nukeq}
\end{equation}
$\nu_0$ is clearly independent of the choice of primitive reciprocal
lattice vectors ${\bf b}_k$ in (\ref{gnk}).  $(\nu_1\nu_2\nu_3)$ are
not.  However, they may be viewed as components of a mod 2 reciprocal
lattice vector,
\begin{equation}
{\bf G}_\nu =
\nu_1 {\bf b}_1 + \nu_2 {\bf b}_2 + \nu_3 {\bf b}_3.
\label{gnueq}
\end{equation}
This vector may be explicitly constructed from the $\delta_i$'s as follows.  A
gauge transformation of the form (\ref{newgauge}) can change the
signs of any four $\delta_i$ for which the $\Gamma_i$ lie in the same
plane.  Such transformations do not change the invariants
$(\nu_1\nu_2\nu_3)$.
By a sequence of these transformations it is always possible
to find a gauge in which $\delta_i = -1$ for at most one nonzero
$\Gamma_i$.  Define $\Gamma^* = \Gamma_i$ if there is one such point.
If there is not, then set $\Gamma^* = 0$.
In this gauge, the mod 2 reciprocal lattice vector is
${\bf G}_\nu = 2\Gamma^*$.  The remaining invariant $\nu_0$ is then
determined by $\delta_i$ at $\Gamma_i=0$.

As we will explain below in section II.C.2, the latter invariants,
$\nu_k$ are not robust in the presence of disorder.  We refer to
them as ``weak" topological invariants. On the other hand, $\nu_0$
is more fundamental, and distinguishes the ``strong" topological
insulator.

The formulas \ref{nueq}-\ref{nukeq} are a bit deceptive because they appear to
depend solely on a  {\it local} property of the wavefunctions.
Knowledge of the {\it global} structure of $|u_{n{\bf k}}\rangle$,
however, is necessary to construct the continuous gauge required
to evaluate Eq. \ref{Z2}.   The existence
of globally continuous wavefunctions is mathematically guaranteed
because the Chern number for the occupied bands vanishes due to
time reversal symmetry. However, determining a continuous gauge
is not always simple.

\subsection{Surface States}

The spectrum of surface (or edge) states as a function of momentum parallel
to the surface (or edge)
is equivalent to the spectrum of discrete end states of the
cylinder as a function of flux.
Fig. 2 schematically shows two possible end state spectra as a
function of momentum (or equivalently flux) along a path connecting
the surface time reversal invariant momenta $\Lambda_a$ and $\Lambda_b$.
Only end states localized at one of the ends of the cylinder is shown.  The shaded
region gives the bulk continuum states.  Time reversal symmetry requires
the end states at $\Lambda_a$ and $\Lambda_b$ be twofold degenerate.
However, there are two possible ways these degenerate states can
connect up with each other.
In Fig. 2a the Kramers pairs ``switch partners" between
$\Lambda_a$ and $\Lambda_b$, while in Fig. 2b they do not.

These two situations are distinguished by the $Z_2$ invariant characterizing
the change in the time
reversal polarization of the cylinder when the flux
is changed between the values corresponding to $\Lambda_a$ and $\Lambda_b$.
Suppose that at the flux corresponding to $\Lambda_a$ the
groundstate is non degenerate, and all levels up to and including
the doublet $\varepsilon_{a1}$ are occupied.  If the flux is adiabatically
changed to $\Lambda_b$ then for Fig 2a the doublet
$\varepsilon_{b1}$ is half filled, and the groundstate has a twofold
Kramers degeneracy associated with the end.  For Fig. 2b, on the other hand,
the groundstate remains non degenerate.  This construction
establishes the connection between the surface states and the bulk
topological invariants.  When $\pi_a\pi_b = -1 (+1)$ the surface spectrum
is like Fig. 2a (2b).

\begin{figure}
 \centerline{ \epsfig{figure=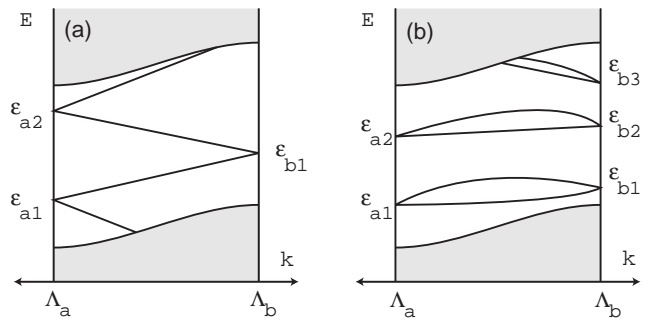,width=3.3in} }
 \caption{Schematic representations of the surface energy levels of
 a crystal in either two or three dimensions as a function of surface
 crystal momentum on a path connecting $\Lambda_a$ and $\Lambda_b$.
The shaded region shows the
 bulk continuum states, and the lines show discrete surface (or edge)
 bands
 localized near one of the surfaces.  The Kramers degenerate surface
 states at $\Lambda_a$ and $\Lambda_b$ can be connected to each other
 in two possible ways, shown in (a) and (b), which reflect the change in time reversal
 polarization $\pi_a\pi_b$ of the cylinder between those points.
 Case (a) occurs in topological insulators, and guarantees the
 surface bands cross any Fermi energy inside the bulk gap.
}
\end{figure}

\begin{figure}
 \centerline{ \epsfig{figure=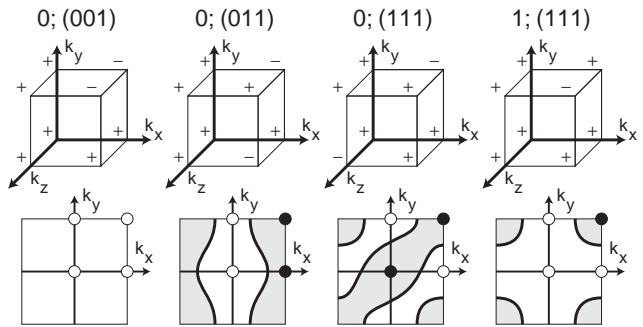,width=3.3in} }
 \caption{  Diagrams depicting four different phases indexed by
 $\nu_0; (\nu_1\nu_2\nu_3)$.  The top panel depicts the signs of
 $\delta_i$ at the points $\Gamma_i$ on the vertices of a cube.  The
 bottom panel characterizes the band structure of a 001 surface for
 each phase.  The solid and open circles depict the time reversal
 polarization $\pi_a$ at the surface momenta $\Lambda_a$,  which
 are projections of pairs of $\Gamma_i$ which differ only in their
 $z$ component.  The
 thick lines indicate possible Fermi arcs which enclose specific
 $\Lambda_a$.
}
\label{phasesfig}
\end{figure}

It follows that when $\pi_a \pi_b=-1 (+1)$ a generic Fermi energy inside
the bulk gap will
intersect an odd (even) number of surface bands between $\Lambda_a$ and
$\Lambda_b$.   Thus, when
$\pi_a\pi_b=-1$ the existence of surface states is topologically
protected.  The details of the surface state spectrum will depend on
the Hamiltonian in the vicinity of the surface.  In Fig. 2 we have
assumed that surface bound states exist for all momenta.  This need
not be the case, since it is possible that by varying the surface
Hamiltonian the degenerate states at $\Lambda_a$ and $\Lambda_b$ can
be pulled out of the gap into the bulk continuum states.  This,
however, does not change our conclusions regarding the number of
Fermi energy crossings.  When $\pi_a\pi_b=-1$ there still must exist
surface band traversing the energy gap.

In the two dimensional quantum spin Hall phase $\pi_1\pi_2=-1$,
and there will be an odd
number of pairs of Fermi points\cite{km1,km2}.  In the simplest case where there is
a single pair, the states at the Fermi energy will be spin filtered in
the sense that the expectation value of the spin in the
right and left moving states will have opposite sign.  These states
are robust in the presence of weak disorder and interactions because time reversal
symmetry forbids elastic backscattering.  Strong interactions,
however, can lead to an electronic instability that opens a gap\cite{wu,xu}.  The
resulting ground state, however, breaks time reversal symmetry.

In three dimensions, the Kramers degenerate band crossings
that occur at $\Lambda_a$ in
the surface spectrum are two dimensional {\it Dirac points}.  While
such Dirac points will occur in any time reversal invariant system
with spin orbit interactions, the nontrivial structure here arises
from the way in which the Dirac points at different $\Lambda_a$ are
connected to each other.  This is determined by the relative signs of
the four $\pi_a$ associated with any surface.

In Fig. 3 we depict four different topological classes for three dimensional
bandstructures labeled according to $\nu_0; (\nu_1\nu_2\nu_3)$, which are
characterized by different values of $\delta_i$ for the
eight $\Gamma_i$, represented as the vertices of a cube in
momentum space.  The lower panel shows the surface Brillouin zone
for a $001$ surface with the four $\Lambda_a$ labeled by either filled or solid circles,
depending on the value of $\pi_a = \delta_{i=(a1)}\delta_{i=(a2)}$.
The surface band structure will resemble Fig. 2(a) on paths
connecting two filled circles or two empty circles, and will resemble
Fig. 2(b) on paths connecting a filled circle to an empty circle.
This allows us to draw a minimal Fermi arc (denoted by the thick lines), which separates regions
containing the filled circles from regions containing the empty
circles.

\subsubsection{Strong Topological Insulator}

For the strong topological insulator, $\pi_1\pi_2\pi_3\pi_4=-1$,
so that a single $\pi_{a=a^*}$ differs in sign from the other
three.  The simplest Fermi arc, shown in Fig 3d thus encloses $\Lambda_{a^*}$ once.
As in the two dimensional case, this Fermi arc can not be
eliminated.  In general, time reversal symmetry requires that any
time reversal invariant Fermi arc $C$ satisfy $C=-C$.
It follows that the Berry's phase acquired by an electron circling
the Fermi arc can only be either 0 or $\pi$.
Since the Fermi arc of the strong topological insulator
encloses a single Dirac point an electron
circling the Fermi arc acquires a Berry's phase of $\pi$.
Since this can not be changed by continuous changes
to the Hamiltonian, we conclude that the $\pi$ Berry's
phase is a generic feature of the surface Fermi arc in the strong
topological insulator phase.  Such a Fermi arc defines a
``topological metal"\cite{fk1}, which is topologically protected
and, unlike an ordinary metal, can not be depleted.

In the presence of weak disorder the $\pi$ Berry's phase changes the
sign of the weak localization correction to the conductivity and
gives rise to antilocalization, as in the symplectic universality
class\cite{suzuura,hikami}.  We suspect that in the absence of
electron-electron interactions these surface states can not be
localized even for strong disorder (provided the bulk topological
phase is not destroyed).  As in the 2D case, however, electron
interactions can lead to a localized phase, which will necessarily
break time reversal symmetry\cite{wu,xu,fk1}.

In the strong topological insulator it is possible that the Fermi energy can
be tuned to intersect a {\it single} Dirac point.  This is a rather
unique situation, because lattice Dirac fermions are generally expected to come
in pairs\cite{nielson}.  These surface Dirac fermions are reminiscent of {\it
domain wall fermions} which have been studied in the context of
lattice gauge theories\cite{kaplan}.  The surface can be viewed as an
interface between the topological insulator and a conventional
insulator (the vacuum).  These two phases can be characterized in
terms of a three dimensional Dirac fermion, whose mass has opposite
sign in the two phases (See for example section III.C).  The domain
wall between the two is then characterized by a gapless Fermion, or zero mode, which is
related to the zero energy midgap states that appear in a one
dimensional Peierls insulator at a soliton\cite{ssh}.
However, there are some important
differences between our model and the conventional applications of domain wall fermions.
(1)  In our problem there is no reason to have
particle-hole symmetry, so tuning is required for the Fermi energy to
be at the Dirac point.  (2)  The domain wall fermion applications
have often been used to model {\it chiral} fermions in ${\it even}$
dimensional space-time\cite{kaplan}.  Our 2+1 dimensional surface Dirac fermions are {\it not}
chiral.  Nonetheless, they realize the 2+1 dimensional ``parity
anomaly"\cite{jackiw}.

The parity anomaly arises for a single (i.e. undoubled)
species of massless Dirac fermion in 2+1 dimensions.  When the response to the
electromagnetic field is naively computed in this model, one
finds\cite{jackiw}
\begin{equation}
J_\lambda = \pm {e^2\over{4 h}} \epsilon_{\mu\nu\lambda} F_{\mu\nu},
\label{jlambda}
\end{equation}
where $J_\lambda$ is the three current and
$F_{\mu\nu}$ is the electromagnetic field tensor in $2+1$ dimensions.  This appears ``anomalous"
in the sense that the electromagnetic field gives rise to currents
which appear to violate the symmetries of the
Dirac Hamiltonian.  The sign ambiguity in (\ref{jlambda}) is due to
the regularization procedure, in which a finite mass is included to
control divergences and taken to zero at the end.
The origin of the singular behavior is the subtlety of this limiting procedure.

In a magnetic field the Dirac equation leads to a Landau level at
exactly zero energy.  At exactly half filling the system is thus at a
critical point separating two quantum Hall states with
$\sigma_{xy}=\pm (1/2)e^2/h$.  This explains the
singular behavior described above.  Indeed, the regulator mass term
discussed above which opens a gap necessarily violates time reversal
symmetry because it lifts a Kramers degeneracy.  This leads to
quantum Hall states even in zero applied magnetic field.

For our problem, in the absence of time reversal symmetry breaking
perturbations we do not expect anomalous currents to occur.  However,
in a magnetic field, the parity anomaly shows up in the
quantum Hall effect because the surface Hall conductivity
will be quantized in {\it half integers},
\begin{equation}
\sigma_{xy} = (n+{1\over 2}){e^2\over h}.
\end{equation}
It is interesting to ask whether such a ``fractional" integer quantum
Hall effect could be measured.  Unfortunately, in a standard
transport experiment in which currents and voltages are measured by
attaching leads to a sample, the $1/2$ can not be directly measured.
The problem is that in a slab geometry there is no simple way to
isolate the ``top" surface from the ``bottom" surface.  The two will
always be measured in parallel, and the two half integers will always
add up to an integer.  For such a transport experiment there is no
getting around the ``fermion doubling problem".  What is required is
a method for measuring the current on the top surface without
measuring the bottom surface.
If it can be established that the
currents are flowing on both surfaces, then dividing the measured Hall
conductance by two could in principle demonstrate the half
quantization.

A lattice realization of the parity anomaly was proposed by Dagotto,
Fradkin and Boyanovski\cite{fradkin1,fradkin2},
who studied a tight binding model for PbTe in the presence
of a domain wall where the Pb and Te atoms are interchanged.  They
showed that in their model the domain wall exhibits massless Dirac fermions,
and has a finite Hall conductivity even at zero magnetic field.
Their model, however is rather different from ours.  In the
presence of the domain wall their Hamiltonian explicitly violates time reversal
symmetry\cite{haldane}, and it leads to an even number of
species of Dirac fermions on the
domain wall.   Haldane introduced a model of the quantum Hall effect on
honeycomb lattice in a periodic magnetic field\cite{haldane}.  This model, which
also breaks time reversal symmetry realizes the parity anomaly
(with a single Dirac fermion) when the Hamiltonian is tuned to
 the transition between the
quantum Hall phase and the insulator.  In this model, however, the
Hall conductivity is an integer.

The surface of the strong
topological insulator is thus unique in that it can generate a single
Dirac fermion {\it without} violating time reversal symmetry, and in principle
exhibits the half quantized quantum Hall effect.

\subsubsection{Weak Topological Insulator}

When $\nu_0=0$, states are classified according to ${\bf G}_\nu$.
We refer to the states with ${\bf G}_\nu \ne 0$ as weak
topological insulators\cite{fkm}.  $\nu_0 = 0$ implies that for any surface
the associated time reversal polarizations will satisfy
$\pi_1\pi_2\pi_3\pi_4 = +1$.  This
implies that either (1) all of the $\pi_a's$ are the same or (2) two will be
positive and two will be negative.  The former case occurs for
surfaces ${\bf G} = {\bf G}_\nu$ mod 2, where ${\bf G}_\nu$ is given
in (\ref{gnueq}).  For these surfaces there are no topologically protected
surface states.  For ${\bf G} \ne {\bf G}_\nu$ mod 2, two of the
$\Lambda_a$'s are positive and two negative.  The
Fermi arc encloses the two $\Lambda_a$'s which have the same sign for $\pi_a$.

These states can be interpreted as layered
two dimensional quantum spin Hall states.
To see this, consider two dimensional planes in the quantum spin Hall
state stacked in the $z$ direction.  When the coupling between the
layers is zero, the electronic states will be independent of $k_z$.
It follows that the four $\delta_i$'s associated with the plane
$k_z=\pi/a$ will have product $-1$ and
 will be the same as the four associated with the plane
$k_z=0$.  The topological invariants will then be given by $\nu_0=0$
and ${\bf G}_\nu = (2\pi/a)\hat z$.  This structure will
remain when weak coupling between the layers is introduced.
More generally, quantum spin
Hall states stacked in the ${\bf G}$ direction will have  ${\bf
G}_\nu = {\bf G}$ mod 2.  This implies that quantum spin Hall states
stacked along different directions ${\bf G}_1$ and ${\bf G}_2$ are
equivalent if ${\bf G}_1={\bf G}_2$ mod 2.

The surface states can also be understood in this manner.  When the
coupling between the layers is zero, it is clear that the gap in the
2D system implies there will be
no surface states on the top and bottom surfaces.  On the sides,
however, the Fermi points guaranteed for the edges of the 2 dimensional system
will become straight Fermi lines, in the $k_z$ direction.  The two
Fermi lines will enclose two time reversal invariant momenta, which
occur at $k_z=0$ and $k_z=\pi/a$, as in Fig. \ref{phasesfig}.

Since the surface Fermi arc encloses an even number of surface time
reversal invariant momenta (and hence an even number of 2D Dirac
points) it follows that the Berry's phase associated with the Fermi
arc is zero.  Thus, the surface states of the
weak topological insulators do not enjoy the
same level of topological protection as those of the strong
topological insulator.  Below we will argue that in the presence of
disorder the weak topological invariants lose their meaning.

\subsection{$Z_2$ invariant and Boundary Condition Sensitivity}

Thouless, Niu, and Wu generalized the topological characterization
of the integer quantum Hall effect to express the Chern invariant
in terms of the sensitivity of the groundstate of a bulk crystal
to phase twisted boundary conditions\cite{thoulessniu}.  This is
more fundamental than the characterization in terms of Bloch
wavefunctions because it does not rely on the translational
symmetry of the crystal.  It explains the topological stability of
the Hall conductance in the presence of weak disorder.  In this
section we consider a corresponding generalization of the $Z_2$
invariant.

To do so, we consider large (but still finite) crystal with
periodic boundary conditions in all but one direction.  A phase
twist $e^{i\theta_i}$ is associated with each periodic boundary
condition. This has the same structure as the cylinder (and
generalized cylinder) considered in section IIB, but now the
circumferences are much larger.  The fluxes
now correspond to the phase twists $\theta_i =\Phi_i e/\hbar$.
Since the cylinder is still finite the discrete states associated
with the ends can be characterized by their degeneracy. This
allows us to characterize the change in time reversal polarization
when the phase twists are changed by $\pi$. For non-interacting
electrons, the invariants characterizing a large cylinder can be
deduced from the bandstructure invariants by treating the entire
sample to be a unit cell of an even larger crystal.  It is
therefore necessary to consider the effect of enlarging the unit
cell on the topological invariants.

The 2D invariant $\nu$ is preserved when the unit cell is enlarged.
This is easiest to see by considering the effect of doubling the unit
cell on the surface spectra of Fig. 2.  The effect of
doubling the unit cell will be to fold the momenta $\Lambda_a$ and
$\Lambda_b$ back on top of each other.  If after enlarging the unit
cell we add a small random potential, which lowers the translational
symmetry, then all ``accidental" degeneracies introduced by the zone
folding will be lifted, while the Kramers degeneracies at the time
reversal invariant momenta will remain.  It is then clear that the
manner in which the degenerate states at $\Lambda_a$ are connected to
each other is preserved when the bands are folded back.  Since the
invariant $\nu$ is unchanged when the unit cell is enlarged, it may
be used to characterized the bulk system.

In three dimensions the strong topological invariant
$\nu_0$ is also invariant under enlarging the unit cell.  This can be
seen by noting that if the surface Fermi arc is folded back, the
number of time reversal invariant momenta enclosed is preserved
modulo 2.  The weak topological invariants $\nu_k$, however, are {\it not}
preserved by this procedure.  It is possible that for a Fermi arc which
encloses two
time reversal invariant momenta the two points can be folded back on top of
each other.  When the two bands are coupled to each other, a gap can
then open at the Fermi energy, so that the Fermi arc can disappear.

We thus conclude that the weak topological invariants are only
defined for the bandstructure of a perfect crystal, and rely on the
lattice translational symmetry.  For a clean system, they
have important implications for the surface state spectrum, but the
topological distinction is eliminated by disorder.  The strong
topological invariant $\nu_0$, however, is robust, and characterizes
a bulk three dimensional phase.

The fragility of the weak topological invariants can also be
understood by considering stacked two dimensional quantum spin Hall
states.  If two identical quantum spin Hall states are coupled
together, the resulting two dimensional system will be a simple
insulator, and will {\it not} have topologcally protected edge
states.  Thus a stack of an even number of layers will be equivalent
to a stack of insulators, while a stack of an odd number of layers
will define a (thicker) 2D quantum spin Hall state.  This sensitivity
to the parity of the number of layers indicates that the weak
topological invariants do not characterize a robust three dimensional phase,
but rather, are only properties of the bandstructure.

This formulation of the $Z_2$ invariant in terms of the change in the
time reversal polarization of a large system as a function of twisted
boundary conditions can be generalized to interacting systems.  This
suggests that the strong topological insulator phase remains robust
in the presence of weak interactions\cite{fk1}.

\section{$Z_2$ Invariants with Inversion Symmetry}

In this section we show how the presence of inversion symmetry
greatly simplifies the evaluation of the $Z_2$ invariants.   We will
prove Eq. 1.1, which allows topological insulators to be identified
in a straightforward manner.

Suppose that the Hamiltonian ${\cal H}$ has an inversion center at
${\bf r}=0$.  It follows that $[{\cal H},P] = 0$, or equivalently
$H(-{\bf k}) = P H({\bf k}) P^{-1}$, where the
parity operator is defined by
\begin{equation}
P|{\bf r}, s_z \rangle = |-{\bf r}, s_z\rangle.
\end{equation}
Here ${\bf r}$ is the (3 dimensional) coordinate and
$s_z$ is the spin which is unchanged by $P$ because
spin is a pseudovector.
The key simplification for problems with combined inversion
and time reversal symmetry is that the Berry
curvature ${\cal F}({\bf k})$
must vanish because it must simultaneously be odd under time reversal
(${\cal F}(-{\bf k})=-{\cal F}({\bf k})$) and even under inversion
(${\cal F}(-{\bf k})=+{\cal F}({\bf k})$)\cite{haldane2}.
The Berry curvature is
given by $\nabla_{\bf k}\times {\cal A}({\bf k})$, where
the Berry's potential is
\begin{equation}
{\cal A}({\bf k}) = -i \sum_{n=1}^{2N} \langle u_{n,{\bf k}}|\nabla_{\bf
k}|u_{n,{\bf k}}\rangle,
\end{equation}
where the sum is over the $2N$ occupied bands.
The fact that  ${\cal F}({\bf k})=0$ suggests it is possible to choose a
globally continuous ``transverse" gauge
in which ${\cal A}({\bf k})=0$.  We will show that in this special
gauge the $\delta_i$ defined in (\ref{Z2})
are given by (1.1), so the $Z_2$ invariants can be
easily evaluated.

The transverse gauge may be explicitly constructed as follows.  In an
arbitrary gauge consider the $2N\times 2N$ matrix
\begin{equation}
v_{mn}({\bf k}) = \langle u_{m,{\bf k}}|P\Theta|u_{n,{\bf k}}\rangle.
\label{vmnofk}
\end{equation}
Since $\langle a|b\rangle = \langle \Theta b|\Theta a\rangle$ and
$\Theta^2 = -1$ it follows that $v({\bf k})$ is
antisymmetric.  Since $[P\Theta,H({\bf k})] = 0$ $v({\bf k})$ is unitary.
Thus the Pfaffian of $v({\bf k})$ is defined, and has unit
magnitude.  The phase of ${\rm Pf}[v({\bf k})]$ depends on the
gauge, and its gradient is related to ${\cal A}({\bf k})$:
\begin{equation}
{\cal A}({\bf k}) = -{i\over 2} {\rm Tr} [ v({\bf k})^\dagger \nabla_{\bf
k} v({\bf k})] =
-i \nabla_{\bf k} {\rm log} {\rm Pf}[v({\bf k})].
\end{equation}
The first equality is derived by differentiating (\ref{vmnofk}) and
using the unitarity of $v({\bf k})$.  The second equality follows
from the facts that ${\rm det}[v] = {\rm Pf}[v]^2$ and
$\nabla_k {\rm log}{\rm det}[v] = {\rm Tr}[\nabla_k {\rm log}\ v({\bf k})] =
{\rm Tr} [ v^\dagger({\bf k}) \nabla_{\bf k} v({\bf k})]$.

To set ${\cal A}({\bf k}) =0$ we thus adjust the phase of
$|u_{n{\bf k}}\rangle$ to make
\begin{equation}
{\rm Pf}[v({\bf k})] = 1.
\end{equation}
This can be done, for instance, by a transformation of the form (\ref{newgauge}),
under which ${\rm Pf}[v({\bf k})] \rightarrow {\rm Pf}[v({\bf k})]
e^{-i\theta_{\bf k}}$.
In this gauge the problem of continuing $\sqrt{{\rm det}[w({\bf
k})]}$ between the $\Gamma_i$ in (\ref{Z2}) is eliminated because
${\rm det}[w({\bf k})] = 1$ for all ${\bf k}$.  This can be seen by
noting that $v({\bf k})$ has the property
$v(-{\bf k}) = w({\bf k}) v({\bf k})^* w({\bf k})^T$ and using the
identity ${\rm Pf}[XAX^T] = {\rm Pf}[A] {\rm det}[X]$.

All that remains for Eq. (\ref{Z2}) is to evaluate ${\rm Pf}[w(\Gamma_i)]$.  To this
end, we note that
\begin{equation}
w_{mn}(\Gamma_i) = \langle \psi_{m,\Gamma_i} |P (P\Theta)|\psi_{n
\Gamma_i}\rangle.
\end{equation}
Here we have used $P^2=1$, along with the anti-linearity of $\Theta$,
which allows us to replace $|u_{n\Gamma_i}\rangle$ by
$|\psi_{n\Gamma_i}\rangle
=|\psi_{n-\Gamma_i}\rangle$ in Eq. (\ref{w}).  Since $[{\cal H},P]=0$,
$|\psi_{n\Gamma_i}\rangle$ is an eigenstate of $P$ with eigenvalue
$\xi_n(\Gamma_i) =\pm 1$.  After changing $|\psi_{n\Gamma_i}\rangle$ back to
$|u_{n\Gamma_i}\rangle$ it follows that
\begin{equation}
w_{mn}(\Gamma_i) = \xi_m(\Gamma_i) v_{mn}(\Gamma_i).
\end{equation}
The Pfaffian can be deduced from the following argument, which uses
the fact that the ${\rm Pf}[w]$ will be a polynomial in $\xi_n$.
First, note that
\begin{equation}
{\rm Pf}[w]^2 = {\rm det}[w] = {\rm det}[v] \prod_{n=1}^{2N} \xi_n.
\label{pfaff2}
\end{equation}
Due the Kramers degeneracy, the distinct states
$|u_{2m,\Gamma_i}\rangle$ and $|u_{2m+1,\Gamma_i}\rangle \equiv
\Theta |u_{2m,\Gamma_i}\rangle$ share the same parity eigenvalue.
Thus, each eigenvalue appears twice in the product in (\ref{pfaff2}).  Taking
the square root, we find
\begin{equation}
{\rm Pf}[w] = {\rm Pf}[v] \prod_{m=1}^N \xi_{2m}
\end{equation}
The sign of the square root is fixed by the special case in which all
$\xi_n=1$, so that $w=v$.  Since ${\rm Pf}[v]=1$
we conclude that in the transverse gauge,
\begin{equation}
\delta_i = \prod_{m=1}^N \xi_{2m}(\Gamma_i).
\label{deltai}
\end{equation}

Eq. (\ref{deltai}) is a central result of this paper.  It means that with
inversion symmetry the $Z_2$
topological invariants can be deduced from the knowledge of the
parity of each pair of Kramers degenerate occupied energy bands at
the four (or eight in 3D) time reversal and parity invariant points in
the Brillouin zone.  This provides a simple method for determining the
topological phase of any inversion symmetric insulator, without
having to know about the global properties of the energy bands.

In Eq. (\ref{deltai}) it appears as though each of the four (or eight) $\delta_i$
have gauge independent meaning, and thus provide extra topological
invariants in addition to the one (or four) time reversal
symmetry based invariants discussed in Section IIB.
These extra invariants, however rely on the presence of
inversion symmetry, and lose their meaning in the presence of
surfaces, disorder or other perturbations which violate inversion
symmetry.  In contrast, the invariants obtained from the
product of 4 $\delta_i$'s do {\it not} rely on inversion symmetry for
their existence.  They depend only on time reversal symmetry, so they
retain their value in the presence of inversion
symmetry breaking perturbations.

\section{Tight-Binding Models}

In this section, we construct a class of
inversion symmetric tight-binding models that exhibit
topological insulating states and apply the method presented
in Section III to determine their topological classes.
We will consider
minimal models with 4 bands which result from four
degrees of freedom per unit cell.  We will focus on
lattices in which the unit cell can be chosen to be inversion
symmetric.  We will see that this latter assumption makes the
analysis of the topological phases particularly simple.
While this assumption can always be satisfied
for continuum models, it rules out certain inversion symmetric
lattice models, such as the rocksalt lattice.  It is satisfied,
however, for the specific examples we will consider.

In IVA we study the general structure of this class of models, and
then in IVB and IVC consider the specific examples of the honeycomb
lattice of graphene and the diamond lattice.  In IVD we analyze a
model for HgTe/CdTe quantum wells introduced recently by Bernevig,
Hughes and Zhang\cite{bhz}.

\subsection{General Model}

We assume each unit cell associated with Bravais lattice vector ${\bf R}$
has four states $|{\bf R},n\rangle$.  If the unit cell is parity
invariant, then the parity operator $P$ may be represented as a $4\times 4$ matrix as
\begin{equation}
P|{\bf R},n\rangle = \sum_m \hat P_{nm} |-{\bf R},m\rangle.
\label{pmnop}
\end{equation}
In sections IVB and IVC we will consider examples in which
each unit cell
consists of two sublattices (denoted by Pauli matrix
$\sigma^z$) which are interchanged by inversion
and two spin degrees of freedom (denoted by $s^z$).  Therefore
\begin{equation}
\hat P = \sigma^x \otimes I,
\label{phatop}
\end{equation}
where $I$ is the identity for the spin indices.  In IVD $\hat P$ will
have a slightly different form.  The time
reversal operator acting on the four component basis states is
represented by
\begin{equation}
\hat \Theta = i (I \otimes s^y ) K,
\label{thetaop}
\end{equation}
where $K$ is complex conjugation, and $I$ acts on the orbital
indices.

Given a lattice Hamiltonian ${\cal H}$ in the $|{\bf R},n\rangle$ basis,
we now consider the Bloch Hamiltonian
\begin{equation}
H({\bf k}) = e^{i{\bf k}\cdot {\bf R}} {\cal H} e^{-i{\bf k}\cdot{\bf
R}},
\label{latticeh}
\end{equation}
which for lattice periodic Bloch functions now becomes a $4\times 4$
matrix.  Note that this transformation is slightly different than
the standard transformation of a tight binding model for which ${\bf R}$
in (\ref{latticeh}) is replaced by ${\bf r} = {\bf R} + {\bf d}_n$
 where ${\bf d}_n$ is a basis vector.  The difference is a choice of
 gauge.  With this choice $H({\bf k})$ has the properties
 $H({\bf k}+{\bf G}) = H({\bf k})$ and $H(-{\bf k}) = \hat P H({\bf
 k}) \hat P^{-1}$.  Thus at the time reversal invariant momenta
 $[H({\bf k}=\Gamma_i),\hat P]=0$.

It is convenient to express the $4 \times 4$ matrix $H({\bf k})$
in terms of the identity $I$, five Dirac matrices
$\Gamma^a$ and their 10 commutators $\Gamma^{ab} =
[\Gamma^a,\Gamma^b]/(2i)$\cite{mnz}.  The Dirac matrices satisfy the Clifford
algebra, $\Gamma^a\Gamma^b+\Gamma^b\Gamma^a = 2\delta_{ab}I$.
In this section,
in order to avoid confusion of notation, the Dirac matrices $\Gamma^a$
will always appear with a superscript, and the time reversal
invariant momenta will always be written as ${\bf k}=\Gamma_i$.

The choice of Dirac matrices is not unique.
For example, in Ref. \onlinecite{km2}, the Dirac
matrices were chosen to be even under time-reversal,
$\hat\Theta\Gamma^a\hat\Theta^{-1} = \Gamma^a$.
In the presence of both inversion and time reversal
symmetry it is more convenient to choose
the Dirac matrices to be even under $\hat P \hat\Theta$.  Given the
form of $\hat P$ and $\hat\Theta$, the five matrices are
\begin{equation}
\Gamma^{(1,2,3,4,5)} = (\sigma^x\otimes
I,\sigma^y\otimes I,\sigma^z \otimes s^x, \sigma^z \otimes s^y,
\sigma^z \otimes s^z).
\label{gamma1-5}
\end{equation}
With this choice of Dirac matrices the commutators are odd under
$\hat P\hat\Theta$, $(\hat P\hat\Theta)\Gamma^{ab}(\hat
P\hat\Theta)^{-1} = -\Gamma^{ab}$.  Note that $\Gamma^1 = \hat P$.
It follows that
\begin{equation}
\hat\Theta\Gamma^a \hat\Theta^{-1} = \hat P \Gamma^a \hat P^{-1} = \left\{
\begin{array}{l}
 +\Gamma^a\quad {\rm for }\ a=1 \\
-\Gamma^a \quad {\rm for }\ a\ne 1.
\end{array}\right.
\label{pandt}
\end{equation}

Time reversal and inversion symmetry imply that $[H({\bf k}),\hat
P\hat\Theta]=0$.  The most general Hamiltonian matrix is then
\begin{equation}
H({\bf k})=d_0({\bf k})I + \sum_{a=1}^5 d_a({\bf k}) \Gamma^a. \label{Gamma}
\label{dirach}
\end{equation}
Written in this form, the energy eigenvalues naturally come in
Kramers degenerate pairs with energy
\begin{equation}
E({\bf k}) = d_0({\bf k}) \pm \sqrt{\sum_a d_a({\bf k})^2}.
\end{equation}

At the time reversal invariant points ${\bf k} = \Gamma_i$, only
$\Gamma^1 = \hat P$ is even under $\hat P$ and $\hat \Theta$.
Therefore
\begin{equation}
H({\bf k} = \Gamma_i) = d_0({\bf k}=\Gamma_i)I + d_1({\bf
k}=\Gamma_i)\hat P.
\end{equation}
The parity eigenvalues $\xi_n$ for the states at
${\bf k}=\Gamma_i$ are given by the eigenvalues of $\hat P$.
It then follows from Eq. \ref{eq1} that provided there is an energy
gap throughout the Brillouin zone, the $Z_2$ invariants characterizing
the valence band are determined by
\begin{equation}
\delta_i=-{\rm sgn} (d_1({\bf k}=\Gamma_i)).
\label{tbdelta}
\end{equation}
We will use the above equation to determine the topological class
of specific tight-binding models in the following.

\subsection{Graphene}

Graphene consists of a honeycomb lattice of carbon atoms with two
sublattices.  A tight-binding model which incorporates the symmetry
allowed spin orbit interactions was introduced in
Refs. \onlinecite{km1} and \onlinecite{km2}.
\begin{eqnarray}
H=t\sum_{\langle ij\rangle} c_i^\dagger c_j +
i \lambda_{SO}
\sum_{\langle\langle ij \rangle\rangle} c_i^\dagger {\bf s} \cdot
 \hat{\bf e}_{ij} c_j . \label{Graphene}
\end{eqnarray}
The first term is a nearest neighbor hopping term,  and
the second term is spin orbit
interaction which involves spin dependent second neighbor hopping.
This term is written in a way which can easily be generalized to
three dimensions.
${\bf s}$ is the spin, and we have defined the unit vector
\begin{equation}
\hat {\bf e}_{ij} =  {  {\bf d}_{ij}^1  \times {\bf d}_{ij}^2\over
{|{\bf d}_{ij}^1  \times {\bf d}_{ij}^2|}},
\end{equation}
where ${\bf d}_{ij}^1$ and ${\bf d}_{ij}^2$ are bond vectors along
the two bonds the electron traverses going from site $j$ to $i$.
Thus, $\hat {\bf e}_{ij} \cdot {\bf s} = \pm s^z$.

\begin{figure}
 \centerline{ \epsfig{figure=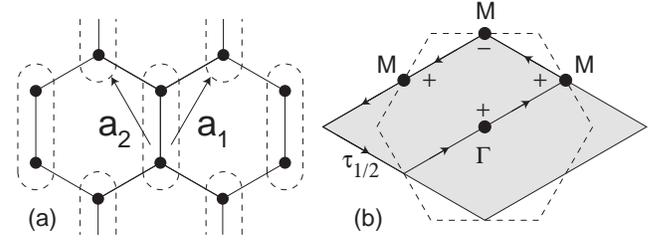,width=3.3in} }
 \caption{ (a) Honeycomb lattice of graphene, with a unit cell
 indicated by the dashed lines.  (b) Brillouin zone, with the values
 of $\delta_i$ associated with the time reversal invariant momenta
 labeled.  $\tau_{1/2}$ describes the loop enclosing half the zone
 used in Eq. \ref{spinless}.
}
\label{graphenefig}
\end{figure}

\begin{table}
  \centering

  \begin{tabular}{|c|c|}
\hline
$d_0$ & $0$  \\
$d_1$ & $t[1+ \cos x_1 + \cos x_2 ]$  \\
$d_2$ & $ t[\sin x_1+\sin x_2]$  \\

$d_3$ & $ 0 $  \\
$d_4$ & $ 0 $  \\
$d_5$ & $2\lambda_{SO} [\sin x_1-\sin x_2-\sin (x_1-x_2)]$  \\
\hline
\end{tabular}
\caption{Parameters for tight-binding model of graphene with
$x_l = {\bf k}\cdot {\bf a}_l$ in a notation
slightly different from Ref. \onlinecite{km2}.}
\label{graphenetab}
\end{table}

Choosing the unit cell shown in Fig. \ref{graphenefig} the
Hamiltonian matrix $H({\bf k})$ can be determined using Eq. \ref{latticeh} and
expressed in terms of Dirac matrices as in \ref{dirach}.
The coefficients $d_a({\bf k})$ are
displayed in Table \ref{graphenetab}.  The time reversal invariant
momenta, in the notation of Eq. \ref{gnk}
occur at ${\bf k}=\Gamma_{i=(n_1n_2)}$, for $n_l=0,1$.
The Hamiltonian at these points can be deduced by noting that
at ${\bf k} = \Gamma_{i=(n_1n_2)}$ $x_l\equiv {\bf
k}\cdot{\bf a}_l = n_l \pi$.   $\Gamma_{i=(00)}$ is
commonly referred to as the $\Gamma$ point.  The other three, which
are equivalent under threefold rotations are called the $M$ points.
Using Eqs. (\ref{eq1},\ref{eq2},\ref{tbdelta}) it is then straightforward to see
that $\delta_{i=(00)}=\delta_{i=(10)}=\delta_{i=(01)} = -1$, while
$\delta_{i=(11)} = +1$.  The product is negative, so $\nu = 1$, and provided the
energy gap is finite throughout the Brillouin zone the system is a
topological insulator in the quantum spin Hall phase.  The finite gap
follows from the fact that $d_0({\bf k})=0$ and there are no values of
${\bf k}$ for which all $d_a({\bf k})=0$.

The reason the three $M$ points are not all
the same is that the center of inversion defined by our unit cell is
at the center of a bond, which does not have the threefold rotational
symmetry.  By choosing a different unit cell, with inversion center
at the center of a hexagon, the $M$ points would be equivalent.  Our
conclusion about the topological class, however, remains the same.

It is interesting to note that the value of $\delta_i$ does not appear
to have anything to do with the spin orbit interaction.  The role
that the spin orbit interaction plays is simply to ensure that the
energy gap is finite everywhere in the Brillouin zone.  We will now
argue for a parity
and time reversal invariant system that
if the spin orbit interaction is {\it absent}, then the
negative product of $\delta_i$ implies that the energy gap {\it must}
vanish somewhere in the Brillouin zone.  This gives insight into the
topological stability of the
Dirac points in graphene in the absence of spin orbit interactions.

We prove this by contradiction.  In the absence of the spin orbit
interaction we can consider spinless fermions.  Suppose there is a
finite gap everywhere, and the valence band is well defined
throughout the Brillouin zone. Then on the one hand, the Berry
curvature ${\cal F} = \nabla\times{\cal A}$ is identically zero due to inversion and time-reversal
symmetry.   On the other hand, we will show that the Berry's phase
for the path $\tau_{1/2}$ shown in Fig. \ref{graphenefig}
which  encloses half the Brillioun zone satisfies
\begin{equation}
e^{ i \oint_{\tau_{1/2}} {\cal A}({\bf k}) \cdot
d {\bf k}} = \delta_1 \delta_2 \delta_3 \delta_4.
\label{spinless}
\end{equation}
Thus if $\delta_1 \delta_2 \delta_3
\delta_4=-1$, it would violate Stoke's theorem and leads to a
contradiction.  The $\pi$ Berry's phase thus requires that there either be
a Dirac point in each half of the Brillouin zone, or a Fermi arc
enclosing a Dirac point.

To obtain Eq.(\ref{spinless}) for spinless electrons, we consider the unitary matrix
\begin{equation}
m_{ij}({\bf k})=\langle u_{i,-{\bf k}} | P | u_{j, {\bf k}}
\rangle,
\end{equation}
which is related to the Berry's potential via
$\nabla_k {\rm log}{\rm det}[m({\bf k})] = -i({\cal A}({\bf k}) + {\cal
A}(-{\bf k}))$.   Eq. (\ref{spinless}) is then
obtained by breaking the line integral into segments connecting
the time reversal invariant momenta and using the fact that
${\rm det}[m({\bf k}=\Gamma_i)]=\delta_i$.

\subsection{Diamond Lattice}

We now consider the tight binding model on a diamond lattice
introduced in Ref. \onlinecite{fkm}.   This model exhibits both weak
and strong topological insulator phases.

The diamond structure consists of two
interpenetrating face-centered cubic lattices displaced from each other
by a basis vector ${\bf d} = a(1,1,1)/4$.
The primitive translation vectors ${{\bf a}_1, {\bf a}_2, {\bf a}_3}$
are $a(0,1,1)/2$, $a(1,0,1)/2$, $a(1,1,0)/2$.
Our model has the same form as Eq. \ref{Graphene}, and includes
a nearest neighbor hopping term as well as a second neighbor spin
orbit interaction.

\begin{table}
  \centering
  \begin{tabular}{|c|c|}
\hline
$d_0$ & $0$  \\
$d_1$ & $t+\delta t_1+t [\cos x_1 + \cos x_2 + \cos x_3]$  \\
$d_2$ & $  t [\sin x_1 + \sin x_2 + \sin x_3]$  \\
$d_3$ & $ \lambda_{SO}[\sin x_2 - \sin x_3 - \sin(x_2-x_1) + \sin(x_3-x_1) ]$  \\
$d_4$ &  $ \lambda_{SO} [\sin x_3 - \sin x_1 - \sin(x_3-x_2) + \sin(x_1-x_2) ] $  \\
$d_5$ & $\lambda_{SO} [\sin x_1 - \sin x_2 - \sin(x_1-x_3) + \sin(x_2-x_3)]$ \\
\hline
\end{tabular}
  \caption{Parameters for diamond lattice tight-binding model,
  with $x_k = {\bf k} \cdot {\bf a}_k$.}
  \label{diamondtab}
\end{table}

It turns out that with this spin-orbit interaction term the
valence bands and conduction bands meet at 3D Dirac points at the
three inequivalent $X$ points on the 100, 010 and 001 faces of the Brillouin zone.
In order to lift the degeneracy and
obtain a gapped phase, we introduced a distortion, which changes the nearest
neighbor hopping amplitudes.  For simplicity we will focus here on a
distortion in the 111 direction, which changes the nearest neighbor
bond in the 111 direction, but leaves the other three bonds alone.
The resulting model can be expressed in the form of Eq.
\ref{dirach}, and the resulting $d_a({\bf k})$ are listed in Table
\ref{diamondtab}.  For $\lambda_{SO}, \delta t \ne 0 $ the gap is finite throughout
the Brillouin zone.

As in the previous section, the time reversal invariant
momenta occur at ${\bf k}=\Gamma_{i=(n_1n_2n_3)}$ as in ({\ref{gnk}).
At these points  $x_l \equiv {\bf
k}\cdot{\bf a}_l = n_l \pi$.
At the $\Gamma$ point ${\bf k}=0$, $(n_1n_2n_3)=(000)$.
The three inequivalent $X$ points (at ${\bf k} = (2\pi/a)(1,0,0)$ and related points)
have $(n_1n_2n_3)=(011),(101)$ and $(110)$.  The four inequivalent $L$ points
(at ${\bf k} = (\pi/a)(1,1,1)$ and related points) have
$(n_1n_2n_3)=(100), (010), (001)$ and $(111)$.  The 111 distortion makes the
first three $L$ points distinct from the fourth, which will be
referred to as $T$.

From Table \ref{diamondtab} we can deduce the sign of $d_1({\bf k})$, and hence
$\delta_i$ at these points.  We find $\delta_i= -1$ at $\Gamma$ and
the three $L$ points, while $\delta_i = +1$ at $T$.  At the $X$ points
$\delta_i=-{\rm sgn}(\delta t_1)$.  Combining these we then find
that
\begin{equation}
(\nu_0;\nu_1\nu_2\nu_3) = \left\{
\begin{array}{ll}
(1; 111) & {\rm for} \quad \delta t_1 >0 \\
(0; 111) & {\rm for} \quad \delta t_1 <0.
\end{array}\right.
\end{equation}
When the 111 distorted bond is stronger than the other three bonds, so that
the system is dimerized, the
system is a strong topological insulator.  When the 111 bond is
weaker than the other three, so that the system is layered, it is a
weak topological insulator with ${\bf G}_\nu = (2\pi/a)(1,1,1)$, which
can be viewed as two dimensional quantum spin Hall states stacked in
the 111 direction.

In Ref. \onlinecite{fkm} we computed the two dimensional band
structure for the diamond lattice model in a slab geometry.  The
results displayed the expected surface states, which behave according
to the general principles discussed in section II.C.

\subsection{Bernevig Hughes Zhang Model}

After this manuscript was originally submitted an interesting
proposal appeared for the 2D quantum spin Hall effect in quantum well
structures in which a layer of HgTe is sandwiched between crystals of
CdTe\cite{bhz}.  Bernevig Hughes and Zhang (BHZ) showed that for an appropriate
range of well thickness, the HgTe layer exhibits an inverted band
structure, where the $s$ and $p$ levels at the conduction and valence
band edges are interchanged.  In this inverted regime, the structure
exhibits a 2D quantum spin Hall effect.  BHZ introduced a simple four
band tight binding model which captures this effect.  Though real
HgTe does not have inversion symmetry, their toy model does.  In this
section we analyze this model and directly evaluate the $Z_2$
topological invariant using (1.1).

BHZ considered a four band model on a square lattice in which each
site has two $s_{1/2}$ states $|s,\uparrow\rangle$, $|s,\downarrow\rangle$
and two of the crystal field split $p_{3/2}$ states (with $m_j = \pm 3/2$),
$|p_x + i p_y,\uparrow\rangle$ and $|p_x-i p_y,\downarrow\rangle$.
The Hamiltonian is
\begin{equation}
H = \sum_{i, \sigma,\alpha}  \varepsilon_\alpha
c_{i \alpha \sigma}^\dagger c_{i \alpha \sigma}
-
\sum_{ia\sigma\alpha\beta} t_{a\sigma,\alpha\beta}
c_{i+a \alpha \sigma}^\dagger c_{i \beta \sigma}
\end{equation}
where $a$ labels the 4 nearest neighbors,
$\sigma=\pm 1$ describes the spin and $\alpha,\beta=s,p$ is
the orbital index.  The hopping term involves the matrix
\begin{equation}
t_{a\sigma} = \left(\begin{array}{cc}  t_{ss}  &
t_{sp}e^{i \sigma \theta_a} \\
t_{sp} e^{-i \sigma\theta_a} & - t_{pp} \end{array} \right)
\end{equation}
where $\theta_a$ gives the angle of nearest neighbor bond
$a$ with the $x$ axis.

As in Section IVA it is convenient to express this Hamiltonian in
the form (\ref{dirach}) using Dirac matrices which are even under
$\hat P \hat \Theta$.  The form of the parity operator, however is
slightly different in this model, and Eq. \ref{phatop} is replaced by
\begin{equation}
\hat P = \sigma^z \otimes I,
\end{equation}
where $\sigma^z = +1 (-1)$ describes $s$ ($p$) states.
The Dirac matrices are then the same as Eq. \ref{gamma1-5}, except
that $\sigma^x$ and $\sigma^z$ are interchanged.  The coefficients of
these new Dirac matrices for this model are displayed in Table III.

\begin{table}
  \centering
  \begin{tabular}{|c|c|}
\hline
$d_0$ & $(\varepsilon_s+\varepsilon_p)/2 - (t_{ss}-t_{pp}) (\cos x_1 + \cos x_2)$  \\
$d_1$ & $(\varepsilon_s-\varepsilon_p)/2 - (t_{ss}+t_{pp}) (\cos x_1 + \cos x_2)$
 \\
$d_2$ & $2 t_{sp}\sin x_1 $  \\
$d_3$ & $  0$  \\
$d_4$ &  $0$  \\
$d_5$ &  $2 t_{sp}\sin x_2$   \\
\hline
\end{tabular}
  \caption{Parameters for the BHZ model,
  with $x_k = {\bf k} \cdot {\bf a}_k$.}
  \label{bhztab}
\end{table}

The analysis between Eqs. \ref{pandt} and \ref{tbdelta} proceeds
exactly the same as before, and $\delta_i = - {\rm sgn}[d_1({\bf
k}=\Gamma_i)]$.  We conclude that for
${\bf k} = (\pi/a)(n_1,n_2)$,
\begin{equation}
\delta_{i=(n_1n_2)} = -{\rm sgn}[ {\varepsilon_s-\varepsilon_p\over 2} -
(t_{ss}+t_{pp})( (-1)^{n_1} + (-1)^{n_2})]
\end{equation}
For $\varepsilon_s - \varepsilon_p >  4(t_{ss}+t_{pp})$ all of the
$\delta_{i=(n_1n_2)}$ are negative, so that the product $\nu=+1$.
The system is a simple insulator.  In this regime the
bands have a conventional ordering throughout the Brillouin zone,
with the $s$ states in the
conduction band and the $p$ states in the valence band.  For
$\varepsilon_s - \varepsilon_p \lesssim  4(t_{ss}+t_{pp})$ the bands
near ${\bf k}=0$ becomes inverted, and $\delta_{i=(00)}$ becomes
positive, signaling a transition into the quantum spin Hall phase in
which $\nu = -1$.

\section{Topological Phases in Specific Materials}

In this section we apply our method for evaluating the topological
invariants to identify specific three dimensional materials that should exhibit a
strong topological insulating phase.

\subsection{Bismuth Antimony Alloy}

Bi and Sb are group V semimetals in which there is a finite direct
energy gap throughout the Brillouin zone, but a negative indirect gap
due to band overlap.  They
have very close lattice parameters and form the solid alloy
${\rm Bi}_{1-x} {\rm Sb}_x$\cite{lerner,lenoir}.  For $.07<x<.22$ the
indirect gap becomes positive, leading to semiconducting behavior,
with a maximum energy gap of order 30 meV for $x = .18$.  In this
section we will argue, based on the known bandstructure of these
materials, that this alloy is a strong topological
insulator, which will have topological metal surface states.

Bulk bismuth and antimony have the rhombohedral A7 structure, which consists
of two interpenetrating, face-centered-cubic
lattices which are displaced in the 111 direction and
slightly distorted in the 111 direction.
In bismuth\cite{golin1,liu}, the
valence band crosses the Fermi energy in the
vicinity of the $T$ point, which is located
on the 111 face of the Brillouin zone, giving rise to a
small pocket of holes.  The conduction band crosses the Fermi energy
near the 3 equivalent $L$ points, which reside at the other three
body center zone faces, giving rise to pockets of electrons.
At the $L$ points, the bottom of the
conduction band, which has $L_s$ symmetry is only slightly higher in
energy than the next lower band, which has $L_a$ symmetry.
In antimony\cite{liu}, the electrons are again near the $L$ point.  However,
unlike bismuth, the bottom of the conduction band has $L_a$ symmetry.
The holes are not at the $T$ point, but rather at the lower symmetry
$H$ point.

Despite the fact that bismuth and antimony have negative indirect
gaps, the finite direct gap throughout the Brillouin zone allows
for the topological characterization of the valence energy bands.
Since both bismuth and antimony have inversion symmetry, we can
apply Eqs. (\ref{eq1},\ref{eq2}) by reading off the parity eigenvalues
$\xi_n(\Gamma_i)$ from published band structures\cite{golin1,liu}.
In Table \ref{bisbtable} we
display the symmetry labels for the five occupied valence bands at
the 8 time reversal invariant momenta ($\Gamma_i = \Gamma$, $T$,
and the three equivalent $L$ and $X$ points). The parity
eigenvalue $\xi_n(\Gamma_i)$ can be read from the superscripts
$\pm$ or the subscripts $s/a=+/-$.  (For an explanation of this
notation see Ref. \onlinecite{falikov}.  The right column displays the
product $\delta_i$ from Eq. (\ref{eq1}). Based on this data, we conclude
that the valence band of bismuth is equivalent to that of a
conventional
insulator, while the valence band of antimony is that of a strong
topological insulator.  The difference between the two is due to
the inversion between the $L_s$ and $L_a$ bands.

\begin{figure}
 \centerline{ \epsfig{figure=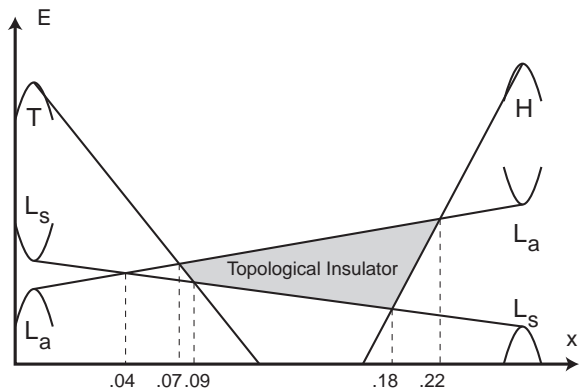,width=3in} }
 \caption{Schematic representation of band energy evolution of ${\rm Bi}_{1-x} {\rm Sb}_x$
 as a function of $x$.  Adapted from Ref. \onlinecite{lenoir}.}
 \label{bisbfig}
 \end{figure}

The evolution of the band structure of the alloy ${\rm Bi}_{1-x} {\rm Sb}_x$ as
a function of $x$ has been well studied
\cite{lerner, golin2,lenoir}, and is summarized in Fig. \ref{bisbfig}.  As the Sb concentration
is increased two things happen.  First, the gap between the $L_s$ and $L_a$
bands decreases.  At $x=.04$ the bands cross and the gap reopens
with the inverted ordering.  Secondly,
the top of the valence band at $T$ comes down in energy
and crosses the bottom of the conduction band at $x=.07$.  At this
point the indirect gap becomes positive, and the alloy is a
semiconductor.   At $x=.09$ the $T$ valence band clears the $L_s$
valence band, and the alloy is a direct gap semiconductor at the $L$
points.
As $x$ is increased further the gap increases until its maximum value
of order 30 meV at $x=.18$.  At that point
the valence band at $H$ crosses the $L_s$ valence
band.  For $x > .22$ the $H$ band crosses the $L_a$
conduction band, and the alloy is again a semimetal.

\begin{table}
  \centering
  \begin{tabular}{|l|c c c c c|c|}
  \multicolumn{7}{c}{\rm Bismuth} \\
 \hline
$1\Gamma$ & $\Gamma_6^+$ & $\Gamma_6^-$ & $\Gamma_6^+$ &
$\Gamma_6^+$ & $\Gamma_{45}^+$ & $-$ \\
\hline
$3L$ & $L_s$ & $L_a$ & $L_s$ & $L_a$ & $L_a$ & $-$\\
\hline
$3X$ & $X_a$ & $X_s$ & $X_s$ & $X_a$ & $X_a$ & $-$\\
\hline
$1T$ & $T_6^-$ & $T_6^+$ & $T_6^-$ & $T_6^+$ & $T_{45}^-$ & $-$ \\
\hline
& \multicolumn{4}{c}{\rm $Z_2$ class}  && $(0;000)$ \\
\hline
\multicolumn{7}{l}{}
\\
  \multicolumn{7}{c}{\rm Antimony} \\
 \hline
$1\Gamma$ & $\Gamma_6^+$ & $\Gamma_6^-$ & $\Gamma_6^+$ &
$\Gamma_6^+$ & $\Gamma_{45}^+$ & $-$ \\
\hline
$3L$ & $L_s$ & $L_a$ & $L_s$ & $L_a$ & $L_s$ & $+$\\
\hline
$3X$ & $X_a$ & $X_s$ & $X_s$ & $X_a$ & $X_a$ & $-$\\
\hline
$1T$ & $T_6^-$ & $T_6^+$ & $T_6^-$ & $T_6^+$ & $T_{45}^-$ & $-$ \\
\hline
& \multicolumn{4}{c}{\rm $Z_2$ class}  && $(1;111)$ \\
\hline

\end{tabular}
  \caption{Symmetry labels for the five valence bands of bismuth and antimony
  at eight time reversal invariant momenta according to Ref. \onlinecite{liu}.  The parity eigenvalues
 can be read from $+/-$ and $s/a$.  Using Eqs (\ref{eq1},\ref{eq2}) they determine
  the topological class.  The indices $(111)$ define a mod 2
  vector (\ref{gnueq}) in the direction of the T point.}
  \label{bisbtable}
\end{table}

Since the inversion transition between the $L_s$ and $L_a$ bands
occurs in the semimetal phase adjacent to pure bismuth, it is clear
that the semiconducting ${\rm Bi}_{1-x} {\rm Sb}_x$ alloy inherits
its topological class from pure antimony, and is thus a strong
topological insulator.  Of course, this conclusion is predicated on a
``virtual crystal approximation" in which the disorder due to the
random mixture is ignored, so that inversion symmetry is preserved in
the alloy.  However, since this inherent disorder does not destroy
the bulk energy gap, it is unlikely to change the topological class,
which does {\it not} require inversion (or translation) symmetry.
We thus conclude that intrinsic ${\rm Bi}_{1-x} {\rm Sb}_x$, despite its bulk
energy gap will have conducting surface states, which form a
topological metal.

Semiconducting ${\rm Bi}_{1-x} {\rm Sb}_x$ alloys have been studied
experimentally because of their thermoelectric properties, which make
them desirable for applications as thermocouples\cite{lenoir,morelli,cho,lin}.
Transport studies
have been carried out both on bulk samples\cite{lenoir} and epitaxial thin films
\cite{cho}.
For $T > 50$K semiconducting behavior is observed, while at lower
temperatures the resistivity saturates at a value in the range
$5-50 \mu\Omega m$.  This observed residual
resistivity is probably too small to be explained by surface states.
It has been attributed to residual charged impurities\cite{lenoir}, which act as
shallow donors, making the alloy slightly $n$ type.  In order to
separate the surface properties from the bulk transport, it will be
necessary either to improve the purity of the samples, or perhaps
use gating in a heterostructure to push the Fermi energy into the
gap.

\subsection{Grey Tin and Mercury Telluride}

Tin is a group IV element, which in it's $\alpha$ (or grey) phase has the
inversion symmetric
diamond structure.  Unlike carbon, silicon and germanium, though, it
is a zero gap semiconductor, in which the ordering of the states at
the conduction and valence band edge is inverted\cite{groves,ngreview}.  The Fermi energy
lies in the middle of a four fold degenerate set of states with
$\Gamma_8^+$ symmetry, which can be derived from $p$ states with total
angular momentum $j=3/2$.  The four fold degeneracy at the $\Gamma_8^+$
point is a consequence of the cubic symmetry of the diamond lattice.
Applying uniaxial strain lifts this degeneracy into a pair of Kramers
doublets and introduces an
energy gap into the spectrum\cite{roman}.  For pressures of order
$3 \times 10^9$ dyn/cm$^2$, the induced energy gap is of order 40 meV.
We now argue that this insulating phase is
in fact a strong topological insulator.

\begin{figure}
\centerline{ \epsfig{figure=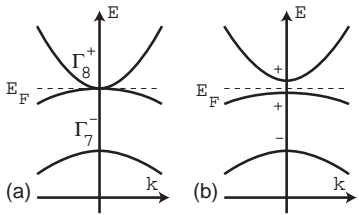,width=2in} }
\caption{(a) Bandstructure of $\alpha$-Sn near the $\Gamma$ point, which describes
zero gap semiconductor due to the inverted $\Gamma_8^+$ and $\Gamma_7^-$ bands.
(b) In the presence of uniaxial strain, the degeneracy at $\Gamma$ is lifted, opening a gap
in the spectrum.  The parity eigenvalues remain unchanged.}
\label{gratinfig}
\end{figure}

\begin{table}
  \centering
  \begin{tabular}{|l|c c c c|c|}
  \multicolumn{6}{c}{\rm Grey Tin } \\
 \hline
$1\Gamma$ & $\Gamma_6^+$ & $\Gamma_7^+$ & $\Gamma_7^-$ &
$\Gamma_8^{+*}$  & $-$ \\
\hline
$3X$ & \multicolumn{2}{c}{$2 X_5$} & $2 X_{5v}$& & $+$\\
\hline
$4L$ & $L_6^-$ & $L_6^+$ & $L_{6v}^-$ & $L_{45}^-$ & $-$\\
\hline
& \multicolumn{3}{r}{\rm $Z_2$ Class} & & $(1;000)$ \\
\hline

\end{tabular}
  \caption{Symmetry labels for the four valence bands of Grey Tin
  at eight time reversal invariant momenta according to Ref.
  \onlinecite{pollak}.}
  \label{tintable}
\end{table}

 Table \ref{tintable} shows the
symmetry labels for unstrained $\alpha-$Sn
associated with the four occupied valence bands at
the eight time reversal invariant momenta\cite{pollak}.  Uniaxial strain lowers
the symmetry, so the cubic symmetry labels no longer apply.
But since the strain does not violate inversion symmetry the
parity eigenvalues are unchanged.  The only effect is to split the
degeneracy of the $\Gamma_8^+$ level into two sets of even parity
 Kramers doublets.
In table \ref{tintable},  $\Gamma_8^{+*}$ refers to the occupied doublet.
Based on the parity eigenvalues we conclude that strained grey
tin is a strong topological insulator.

HgTe is a II-VI material with the zincblend structure\cite{ngreview,hgtereview}.
It is a zero gap semiconductor
with an electronic structure closely related to grey tin.
The Fermi energy is in the middle of the four fold degenerate
$\Gamma_8$ states, whose degeneracy follows from the cubic symmetry
of the zincblend lattice.  As in grey tin, uniaxial strain lifts this
degeneracy and opens a gap at the Fermi energy.

Though HgTe lacks inversion symmetry, we now argue based on adiabatic
continuity that the gap induced by uniaxial strain leads to a strong
topological insulator.  The electronic structure of II-VI materials
can be understood by adding an inversion symmetry breaking
perturbation to a inversion symmetric group IV crystal\cite{ngreview,herman}.
Provided
this perturbation does not lead to any level crossings at the Fermi
energy, we can conclude that the II-IV material is in the same
topological class as the group IV crystal.
The bandstructures of grey tin and HgTe are very similar, and the
cubic symmetry labels of the energy bands show how the bands evolve
between the two.  This allows us to conclude that strained
$\alpha$-Sn and HgTe will be in the same topological class, which is
that of the strong topological insulator.

In Ref. \onlinecite{murakami2} Murakami, Nagaosa and Zhang introduced
a four band tight binding model based on $p_{3/2}$ atomic levels on a fcc
lattice to describe strained $\alpha$-Sn and HgTe.
As argued in Ref. \onlinecite{km2}, this model predicts that these
materials are simple insulators in the $0;(000)$ class.  This can be
understood by noting that since the model includes only $p_{3/2}$ atomic
levels the parity eigenvalues in (\ref{eq1}) are all $\xi_i=-1$.  This
contradicts the known band structure of these materials, as displayed
in Table \ref{tintable}.  This model correctly describes the
electronic states near the $\Gamma$ point, but it gets the global
topology of the bands wrong.  To capture the global topology a tight
binding model of these materials must include both $s$ and $p$ levels.
The more recent theory\cite{bhz} of the 2D quantum spin Hall effect
in HgTe/CdTe quantum wells discussed in
Section IVD correctly incorporates $s$ and $p$ levels.

\subsection{Lead-Tin Telluride}

PbTe and SnTe are narrow gap IV-VI semiconductors with the rocksalt
structure\cite{pbtereview}.   The bandgap in
these materials is direct, and occurs at the 4 equivalent $L$ points in the FCC
Brillouin zone.  PbTe has an inverted bandstructure relative to SnTe,
in which the relative ordering of the $L_6^+$ and $L_6^-$ bands at the
conduction and valence band edges are switched.  Nonetheless, both of
these materials are conventional insulators.  In Table \ref{pbtetable}, we display
the symmetry labels at the 8 time reversal invariant points
($\Gamma$, 3 equivalent $X$ points and 4 equivalent $L$ points)\cite{tung}.
Since the inversion occurs at an even number of points in the
Brillouin zone, both materials belong to the conventional insulator
topological class.

PbTe and SnTe form an alloy Pb$_{1-x}$Sn$_x$Te.  At $x\sim .4$ there is
an inversion transition where the band gap at the four $L$ points
vanishes, giving rise to three dimensional Dirac points\cite{pbtereview,dimmock}.
The phases on either side of this transition are only distinct if
inversion symmetry is present.  Thus disorder, which is inevitably
present in the alloy blurs the transition.  However, uniaxial strain
oriented along the 111 direction will distinguish one of the $L$
points (call it $T$ now) from the other three $L$ points.  It follows that
the inversion transitions at the $L$ and $T$ points will occur at
different values of $x$.  Thus there will be an intermediate phase in
which $L$ is inverted, but $T$ is not (or vice versa).  From Eqs.
(1.1,1.2) this intermediate phase will be a strong topological
insulator with surface states forming a topological metal.
Note, that this direction depends on the {\it orientation} of the
uniaxial strain.  For instance strain along the 100 direction will
distinguish {\it two} $L$ points from the other two, and will {\it
not} lead to an intermediate topological phase.
In Fig. 2 we show a schematic phase diagram as a function of $x$ and
111 strain.

\begin{table}
  \centering
  \begin{tabular}{|l|c c c c c|c|}
  \multicolumn{7}{c}{\rm Tin Telluride} \\
 \hline
$1\Gamma$ & $\Gamma_6^+$ & $\Gamma_6^+$ & $\Gamma_6^-$ &
$2\Gamma_8^+$ &  & $-$ \\
\hline
$3X$ & $X_6^+$ & $X_6^+$ & $X_6^-$ & $X_6^-$ & $X_7^-$ & $-$\\
\hline
$4L$ & $L_6^-$ & $L_6^+$ & $L_6^+$ & $L_{45}^+$ & $L_6^-$ & $+$\\
\hline
& \multicolumn{4}{c}{\rm $Z_2$ class}  && $(0;000)$ \\
\hline
\multicolumn{7}{l}{}
\\
  \multicolumn{7}{c}{\rm Lead Telluride} \\
 \hline
$1\Gamma$ & $\Gamma_6^+$ & $\Gamma_6^+$ & $\Gamma_6^-$ &
$2\Gamma_8^+$ &  & $-$ \\
\hline
$3X$ & $X_6^+$ & $X_6^+$ & $X_6^-$ & $X_6^-$ & $X_7^-$ & $-$\\
\hline
$4L$ & $L_6^-$ & $L_6^+$ & $L_6^+$ & $L_{45}^+$ & $L_6^+$ & $-$\\
\hline
& \multicolumn{4}{c}{\rm $Z_2$ class}  && $(0;000)$ \\
\hline

\end{tabular}
  \caption{Symmetry labels for the five valence bands of Tin
  Telluride and Lead Telluride at eight time reversal invariant momenta,
  according to Ref. \onlinecite{tung}}
  \label{pbtetable}
\end{table}

\begin{figure}
 \centerline{ \epsfig{figure=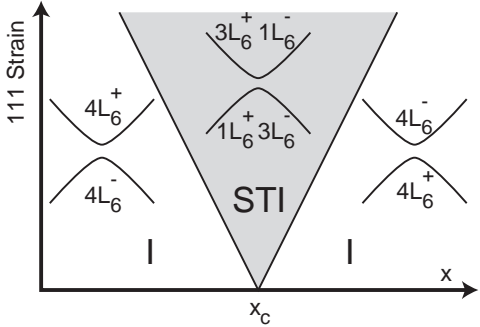,width=2.5in} }
 \caption{Schematic uniaxial strain-composition phase diagram for Pb$_{1-x}$Sn$_x$Te.
 Away from the inversion transition at $x\sim .4$ the material is a conventional
 insulator (I).  Near the transition it is a strong topological insulator (STI). }
 \end{figure}

The band inversion between SnTe and  PbTe has been discussed by a number of
authors previously.  Volkov and Pankratov\cite{volkov} argued that
 PbTe and SnTe can be described by a low energy
field theory consisting of three dimensional Dirac fermions with masses
of opposite sign.  They concluded that a planar interface between PbTe
and SnTe will exhibit interface states described by a
two dimensional massless Dirac equation.  The appearance of such
``domain wall fermions" is similar to the appearance of midgap states
in a one dimensional Peierls insulator at a soliton\cite{ssh}.
A related proposal was made
by Fradkin, Dagotto and Boyanovski\cite{fradkin1,fradkin2}, who considered a domain wall in
PbTe in which on one side the Pb and Te atoms are interchanged.  This
was also modeled as 3D Dirac fermions with a mass which changes sign at the
interface.

The domain wall fermions which appear in these theories are
similar to the states we predict at the surface of
a strong topological insulator.  Indeed, if one views the vacuum as a
band insulator with a large gap, then the surface can be viewed as an
interface between a band insulator and a topological insulator, which
can be described as an inversion transition,
where there is a change in the sign of the mass of a 3D
Dirac fermion.
However, there is an important difference between the proposals
discussed above and the surface states of the topological insulator:
the strong topological insulator - band insulator interface involves
a sign change in an {\it odd} number of Dirac points, while the
interface models above involve {\it four} Dirac points.  Having an odd number
is crucial for the topological stability of the surface states.

\subsection{Other materials}

The materials we have proposed above should not be considered to be
an exhaustive list.  In general it is necessary to consider
insulators composed of heavy elements.  Another candidate for a
topological insulating phase is Bi$_2$Te$_3$, which, like
Bi$_{1-x}$Sb$_x$, is known for its thermoelectric properties\cite{mishra}.  This
material is also a narrow gap semiconductor, with an energy gap of
order .13 eV.  Though the crystal structure of this material
is inversion symmetric, we have been unable to locate band theory
calculations which display the parity eigenvalues.

Another possible candidate is the zincblend semiconductor
$\beta$-HgS.  The electronic structure of this material has been a subject
of some controversy.  According to Delin\cite{delin}, it is a semiconductor
which has an unusual band ordering, with the $\Gamma_6$ and $\Gamma_8$
levels in the valence band and the $\Gamma_7$ level in the conduction band.
If this is the case, we expect the material to be a strong topological insulator.
However, this conclusion has
been challenged by Moon, et al. \cite{moon}, who find a more
conventional band ordering with the $\Gamma_6$ level in the
conduction band and the $\Gamma_7$ and $\Gamma_8$ levels in the
valence band, leading to a conventional insulator.

\section{Experimental Implications}

We now briefly consider possible experimental probes of topological
insulators.  We will focus here on the three dimensional strong
topological insulator phase, for which we suggested several materials
in the previous section.

The most direct probe of the topological insulators is transport.
Since there is a bulk excitation gap, transport in intrinsic samples
at very low temperature will be dominated by the surfaces, which can
be probed by the geometry dependence of the conductance.  For
example, for a wire geometry the conductance will be proportional to
the circumference of the wire, rather than the area.

In addition, since the topological metal phase is in the symplectic
universality class the conductivity is expected to increase logarithmically
at low temperature: $\sigma(T) \propto {\rm log}
[L_{\rm in}(T)/\ell]$\cite{senthil}, where $L_{\rm in}$ is the inelastic scattering
length and $\ell$ is the mean free path.

An interesting prediction for the surface states is that due the
$\pi$ Berry's phase associated with the surface Fermi arc, the
surface quantum Hall effect in a perpendicular magnetic field should be
quantized in half odd integers, $\sigma_{xy} = (n+1/2)e^2/h$.
As discussed in section II.C.1 this is  difficult
to measure directly without separately measuring the currents flowing on the
top and bottom surfaces of the sample.  However if the parallel
combination of the two surfaces could be measured, the resulting Hall
effect would be quantized in {\it odd} multiples of $e^2/h$.  This is similar to
the quantum Hall effect in graphene\cite{geim,kim}, which is quantized in odd
multiples of $2e^2/h$.  The difference is due to the fact that
graphene has four Dirac points, including spin.

A practical difficulty with transport measurements is that it is
necessary to distinguish surface currents from bulk currents.
Since bulk currents scale with the sample width $W$, even though there is
a bulk energy gap $E_g$, the temperature
must be low: $T \ll E_g /{\rm log} W/a$, where
 $a$ is the lattice constant.   Moreover, since the materials we
have suggested have rather small energy gaps, samples with high
purity will be required to reach the intrinsic limit.  As discussed
in section IVa, the low temperature behavior of Bi$_{1-x}$ Sb$_x$
is dominated by a low concentration of charged impurities, which form
an impurity band\cite{lenoir}.  This is a ubiquitous problem for narrow gap
semiconductors, due to their low effective mass and high dielectric
constant.  Clearly it would be desirable to have a transport geometry which
probes the surface states, while controlling the position of the bulk
Fermi energy.  Perhaps this may be possible in a clever
heterostructure geometry, where the bulk Fermi energy can be adjusted
with a gate.

An alternative probe of the surface states would be to map the
surface Fermi arc using angle resolved photo emission.  Such
measurements could establish that the surface Fermi arc
encloses an odd number of time reversal invariant momenta in the
strong topological insulator phase.  Detailed
ARPES studies have been carried out on the surfaces of bismuth\cite{ast, gayone, hofmann}
and antimony\cite{sugawara}.  However, the presence of the bulk Fermi
surface complicates the analysis of these materials.
It would be interesting to see how the results are modified in the semiconducting Bi$_{1-x}$
Sb$_x$ alloy.

Finally, since the surface states are spin filtered, electrical
currents flowing on the surface will be associated with spin
accumulation, leading to a spin Hall effect.  In GaAs, spin accumulation
on a surface has been measured\cite{kato,wunderlich}.
The narrow energy gaps in
our proposed materials make detection of the spin accumulation
more difficult.  Perhaps a heterostructure geometry could make this
possible.

\section{Conclusion}

In this paper we discussed topological insulator phases in two
and three dimensions.  We discussed in detail how the $Z_2$ topological
invariants characterizing these phases influence the surface state
spectrum.  In particular, the quantum spin Hall phase in two
dimensions and the strong topological insulator in three dimensions
have robust conducting surface states, which we have characterized as
a topological metal.  We showed that the $Z_2$ invariants can be
determined easily in parity invariant crystals from the knowledge of
the parity eigenvalues for states at the time reversal invariant
points in the Brillouin zone.  Using this method, we deduced that the
semiconducting alloy Bi$_{1-x}$Sb$_x$ is a strong topological
insulator, as are $\alpha$-Sn and HgTe in the presence of uniaxial
strain.

There remain a number of further issues which need to be understood better.
High among them are the effects of disorder and
interactions.  These are important both for the topological metal
surface states as well as for the bulk topological phases.
Numerical work by Onoda et al. \cite{onoda} has
suggested that the transition between the conventional insulator and
the quantum spin Hall phase in two dimensions belongs to a new
universality class.  It will be of interest to understand this
transition better, along with the related transition between the topological
insulator and the Anderson insulator, which presumably occurs when
 disorder is increased beyond a critical value.

Finally, it would be desirable to develop a field theory for the
topological insulating phases analogous to the Chern Simons theory of
the quantum Hall effect.  Perhaps this may lead to analogs of the fractional
quantum Hall effect for the topological insulators.

\acknowledgments

It is a pleasure to thank Eduardo Fradkin and Gene Mele for
helpful discussions.   This work was supported by NSF grant
DMR-0605066, and by ACS PRF grant 44776-AC10.

\end{document}